# RC100: Rotation Curves of 100 Massive Star-Forming Galaxies at z=0.6-2.5 Reveal Little Dark Matter on Galactic Scales

A. Nestor Shachar[1], S.H. Price[2], N.M. Förster Schreiber[2], R. Genzel[2,3], T.T. Shimizu[2], L.J. Tacconi[2], H. Übler[11,12], A. Burkert[4], R.I. Davies[2], A. Dekel[6], R. Herrera-Camus[7], L. L. Lee[2], D. Liu[2], D. Lutz[2], T. Naab[8], R. Neri[5], A. Renzini[9], R. Saglia[3], K. Schuster[4], A. Sternberg[1,2,10], E. Wisnioski[13,14] & S. Wuyts[15]

[1]*School of Physics and Astronomy, Tel Aviv University, Tel Aviv 69978, Israel (amitnestor@mail.tau.ac.il)*
[2]*Max-Planck-Institut für Extraterrestrische Physik (MPE), Giessenbachstr.1, 85748 Garching, Germany (genzel@mpe.mpg.de, forster@mpe.mpg.de, linda@mpe.mpg.de)*
[3]*Departments of Physics and Astronomy, University of California, 94720 Berkeley, USA*
[4]*Universitäts-Sternwarte Ludwig-Maximilians-Universität (USM), Scheinerstr. 1, München, D-81679, Germany (burkert@usm.uni-muenchen.de)*
[5]*Institute for Radio Astronomy in the Millimeter Range (IRAM), Rue de la Piscine, Grenoble, France*
[6]*Racah Institute of Physics, The Hebrew University of Jerusalem, Jerusalem 9190401, Israel*
[7]*Astronomy Department, Universidad de Concepción, Av. Esteban Iturra s/n Barrio Universitario, Casilla 160, Concepción, Chile*
[8]*Max-Planck Institute for Astrophysics, Karl Schwarzschildstrasse 1, D-85748 Garching, Germany*
[9]*Osservatorio Astronomico di Padova, Vicolo dell'Osservatorio 5, Padova, I-35122, Italy*
[10]*Center for Computational Astrophysics, Flatiron Institute. 162 5th Avenue, New York, NY, 10010, USA*
[11]*Kavli Institute for Cosmology, University of Cambridge, Madingley Road, Cambridge, CB3 0HA, UK*
[12]*Cavendish Laboratory - Astrophysics Group, University of Cambridge, 19 JJ Thompson Avenue, Cambridge, CB3 0HE, UK*
[13]*Research School of Astronomy and Astrophysics, Australian National University, Canberra, ACT 2611, Australia*
[14]*ARC Centre of Excellence for All Sky Astrophysics in 3 Dimensions (ASTRO 3D)*
[15]*Department of Physics, University of Bath, Claverton Down, Bath, BA2 7AY, United Kingdom*

# Abstract

We analyze Hα or CO rotation curves (RCs) extending out to several galaxy effective radii for 100 massive, large, star-forming disk galaxies (SFGs) across the peak of cosmic galaxy star formation ($z \sim 0.6 - 2.5$), more than doubling the previous sample presented by Genzel et al. (2020) and Price et al. (2021). The observations were taken with SINFONI and KMOS integral-field spectrographs at ESO-VLT, LUCI at LBT, NOEMA at IRAM, and ALMA. We fit the major axis kinematics with beam-convolved, forward models of turbulent rotating disks with bulges embedded in dark matter (DM) halos, including the effects of pressure support. The fraction of dark to total matter within the disk effective radius ($R_e \sim 5$ kpc), $f_{DM}(R_e) = V_{DM}^2(R_e)/V_{circ}^2(R_e)$, decreases with redshift: At z~1 (z~2) the median DM fraction is $0.38 \pm 0.23$ ($0.27 \pm 0.18$), and a third (half) of all galaxies are "maximal" disks with $f_{DM}(R_e) < 0.28$. Dark matter fractions correlate inversely with the baryonic surface density, and the low DM fractions require a flattened, or cored, inner DM density distribution. At $z \sim 2$ there is $\approx 40\%$ less dark matter mass on average within $R_e$ compared to expected values based on cosmological stellar-mass halo-mass relations. The DM deficit is more evident at high star formation rate (SFR) surface densities ($\Sigma_{SFR} \gtrsim 2.5 \, M_\odot \, yr^{-1} \, kpc^{-2}$) and galaxies with massive bulges ($M_{bulge} \geq 10^{10} \, M_\odot$). A combination of stellar or active galactic nucleus (AGN) feedback, and/or heating due to dynamical friction, either from satellite accretion or clump migration, may drive the DM from cuspy into cored mass distributions. The observations plausibly indicate an efficient build-up of massive bulges and central black holes at $z\sim 2$ SFGs.

# 1. Introduction

*Low dark matter fractions at z~2.* Galaxy rotation curves (RCs) are an essential tool for studying the baryonic and dark matter (DM) mass distributions of galaxies. Several studies in the past decade have used advanced optical and near-IR imaging spectrographs and millimeter interferometry to obtain RCs for high redshift star forming galaxies (SFGs) (e.g., Wuyts et al. 2016; Stott et al. 2016, Genzel et al. 2017, 2020; Lang et al. 2017; Übler et al. 2018; Tiley et al. 2019; Price et al. 2021; Bouché et al. 2021). Deep observations and stacking techniques are critical to probe RCs beyond the disk effective radii, $R_e$, out to where the RC shapes are more sensitive to the presence of dark matter. Genzel et al. (2017; hereafter G17) analyzed high S/N Hα kinematics taken from the KMOS$^{3D}$ and SINS/zC-SINF surveys (Wisnioski et al. 2015, 2019; Förster Schreiber et al. 2009, 2018) for six individual massive disks at $z = 0.9 - 2.4$, reaching ~2-3 $R_e$. Detailed forward mass modeling of the major-axis rotation velocity and velocity dispersion profiles indicated that ***all six galaxies are strongly baryon-dominated within $R_e$ having very low derived DM fractions***, $f_{DM}(R_e) = V_{DM}^2(R_e)/V_{circ}^2(R_e) < 21\%$. Lang et al. (2017) stacked RCs for ~100 massive, main-sequence SFGs from the KMOS$^{3D}$ and SINS/zC-SINF surveys, finding significant decline in the circular velocity at large radii. Genzel et al. (2020; hereafter G20) analyzed individual RCs extending to several $R_e$ of 35 additional massive disks (hereafter, RC41) taken from the KMOS$^{3D}$ and SINS/zC-SINF surveys, as well as sensitive IRAM/NOEMA CO data sets of $z \sim 1$ SFGs (Tacconi et al. 2013, 2018; Freundlich et al. 2019). This larger sample showed that at $z \sim 2$ disk galaxies have very low DM fractions ($f_{DM}(R_e) < 20\%$), and revealed a strong anti-correlation between the DM fractions and baryonic mass and averaged baryonic mass surface densities, $\Sigma_{bar}(<R_e)$, In a companion paper, Price et al. (2021) demonstrated that the G20 results are little affected by the fitting methodology, and that fitting the 2D velocity field yields similar results as major-axis 1D RC fitting and indicating that ***most of the global dynamical information is captured along the major-axis***. Other studies of SFGs at high-z have analyzed RCs based on Hα or CO observations of various samples, in broad agreement with the results obtained in G20 when considering comparable aperture radii (Price et al. 2016, 2020; Wuyts et al. 2016; Lang et al. 2017; Drew et al. 2018; Molina et al. 2019; Tiley et al. 2019; Bouché et al. 2021).

*Dark matter cores.* Very low DM fractions suggest flat ("cored") DM mass distributions in the central regions of the halo, in contrast with expectations from cosmological simulations predicting DM mass distribution that diverge towards the center ("cusps", e.g., NFW, see Moore et al. 1999; Klypin et al. 2001; Power et al. 2003; Navarro et al. 2004, 2010; Dekel et al. 2017). This difference is amplified if one considers adiabatic contraction of the halo by its baryonic content, resulting in steep inner mass distributions (see Blumenthal, Flores & Primack 1986; Mo, Mao & White 1998). The observed low DM fractions at $z\sim 2$ appear more consistent with cores at the scale of the disk effective radius, $R_e$. In the Local Universe, DM cores have been seen extensively in dwarf galaxies but not for massive disks (see review by Boldrini 2021, also Spekkens et al. 2005; Kuzio de Naray et al. 2006; Faerman et al. 2013, Oh et al. 2015; Governato et al. 2012; Cooke et al. 2022). For cores to form in massive galaxies, adiabatic contraction needs to be countered by opposing mechanisms. Stellar and active galactic nuclei (AGN) feedback can become effective at high star-formation and black hole accretion rates, redistributing the DM on the galactic scale and heating via dynamical friction and small satellite mergers may cause the DM to be more reactive to feedback (Navarro Frenk & White 1996; El Zant et al. 2001; Ogiya et al. 2011, 2022; De Souza et al. 2011; Chan et al. 2015; Schulze et al. 2020; Dekel et al. 2021). In the FIRE simulation, detailed stellar feedback processes form DM cores in massive dwarf galaxies at $z \sim 0$, but they still result in cusps for more massive disk galaxies (Lazar et al. 2018).

*Advantages of high-z RCs.* At higher redshifts, the galactic disk and the DM halo are less prone to degeneracies in their contribution to the total RC. The small angular sizes of high-z galaxies make determinations of their dynamics much more challenging than for low-z disks, but they are most useful in constraining the DM content. At $z \sim 1-2$ a typical disk galaxy is smaller by a factor 2-3 than a galaxy with similar virial velocity in the Local Universe (for $\Lambda$=0.7, Mo, Mao & White 1998). Meanwhile, the concentration parameter ($c = r_{vir}/r_s$) of the DM halo is predicted to be also smaller by a factor 2-3. (Navarro, Frenk & White 1996; Bullock et al. 2001; Dutton & Maccio 2014; Ludlow et al. 2014), mainly due to the smaller virial radii at higher redshift. The (inner) scale radii of the DM halos remain fairly constant. The disk and halo components are therefore better distinguishable at higher z, reducing the degeneracy between disk and halo (van Albada & Sancisi 1986). Furthermore, velocity dispersions of disks at $z \sim 2$ are higher by a factor 2-3 than their Local Universe counterparts o (e.g., Kassin et al. 2006; Förster Schreiber et al. 2006; Épinat et al. 2012; Wisnioski et al. 2015; Stott et al. 2016; Simons et al. 2017; Übler et al. 2019). These high velocity dispersions create pressure gradients in the disk, leading to partial support and a radial drop in the rotation velocity ("asymmetric drift correction", see Burkert et al. 2010, 2016). Crucially, if sizeable, this drop is direct evidence that the dark matter contributions to the RC are low and the baryons dominate the gravitational potential.

*This work.* We expand on Genzel et al. (2017, 2020) and Price et al. (2021) and present "RC100": our third generation set of outer disk RCs from deep observations of Hα and/or CO. We add another 59 galaxies for a total of 100 high-quality extended RCs at a redshift span between z = 0.6 – 2.5, increasing statistics and extending the galaxy parameter coverage to smaller radii across the star-formation MS. The galaxies sample the main sequence of star formation, with specific characteristics as described in Section 2. Each galaxy is forward modelled using three different fitting procedures, from which the galaxy properties are extracted. We combine the best-fit values from all three methods, finding little effect of systematic differences arising from fitting procedures.

In Section 2 we present the RC100 sample properties, in Section 3 the fitting methodologies used, and in Section 4 the main results of the analysis: the dependence of $f_{DM}(R_e)$ on the circular velocity and baryonic surface density, as well as the amount of inferred dark matter, its "deficit", and its connection to the star formation rate (SFR) surface density. In section 5 we discuss and summarize our results. Throughout the paper, we assume a flat $\Lambda CDM$ cosmology with $\Omega_m = 0.3, \Omega_\Lambda = 0.7$ and $H_0 = 70 \text{ km s}^{-1} Mpc^{-1}$, and a Chabrier (2003) stellar initial mass function.

## 2. RC100 Sample and Data Sets

*Instruments and surveys.* The RC100 sample, as for the RC41 subset (See G20), is mostly drawn from two near-IR Integral Field Unit (IFU) spectroscopic samples with deep observations of Hα kinematics carried out at the Very Large Telescope (VLT), totaling ∼800 SFGs: the seeing-limited KMOS[3D] survey[1] with the KMOS multi-IFU at $0.6 < z < 2.6$ (Wisnioski et al. 2015, 2019), and the SINS/zC-SINF survey[2] at $1.3 < z < 2.6$ with adaptive-optics (AO) and seeing-limited SINFONI data (Förster Schreiber et al. 2009; 2018; Mancini et al. 2011). In addition, we include targets with resolved kinematics from mm CO line emission observed with the IRAM NOEMA interferometer in the PHIBSS and NOEMA[3D] surveys[3] (Tacconi et al. 2013, 2018; Freundlich et al. 2019; NOEMA[3D] team, in prep.), two of which also have long-slit resolved major-axis Hα spectroscopy obtained with the LUCI spectrograph at the Large Binocular Telescope (LBT; Genzel et al. 2013; Übler et al. 2018). Finally, we include 3 targets with new high-resolution ALMA CO data (see also Liu et al., in prep., for J0901+1814). Table 1 summarizes the number of targets obtained from the different telescopes with the corresponding emission lines. The total on-source integration time is 1418 hours, with a median of 10.7 hours per galaxy. 10 SFGs have integration times over 25 hours. 34 SFGs have data from more than one instrument and five combine both Hα and CO data. For these latter cases the agreement in the derived rotation curves and mass distributions is excellent (Genzel et al. 2013, Übler et al. 2018, Davies et al. 2020).

| Instrument (line) | Number of galaxies | Reference |
|---|---|---|
| SINFONI seeing limited (Hα) | 34 | Förster Schreiber et al. 2009; Mancini et al. 2011 |
| SINFONI-AO (Hα) | 22 | Förster Schreiber et al. 2018 |
| KMOS seeing limited (Hα) | 64 | Wisnioski et al. 2015, 2019 |
| LBT (Hα) | 2 | Genzel et al. 2013; Übler et al. 2018 |
| NOEMA (CO) | 9 | Tacconi et al. 2013; 2018; Freundlich et al. 2019, NOEMA[3D] team, in prep. |
| ALMA (CO) | 3 | Liu et al., in prep |

*Table 1: Number of RC100 galaxies observed by each instrument with the corresponding observed line. 34 Galaxies have been observed with more than one instrument, and 5 galaxies combine both Hα and CO observations.*

---

[1] The fully reduced KMOS data cubes and catalog of galaxy properties are publicly available at https://www.mpe.mpg.de/ir/KMOS3D/data.
[2] The fully reduced SINFONI data cubes are publicly available at https://www.mpe.mpg.de/ir/SINS/SINS-zcSINF-data.
[3] https://www.iram-institute.org/EN/content-page-279-7-158-240-279-0.html and https://www.iram-institute.org/EN/content-page-419-7-57-412-415-418.html.

All details of the target selection, sample properties (including stellar mass, SFR, and size estimates), observations, and data reduction of the RC100 parent Hα or CO surveys are given in the references in Table 1. In brief, the targets were primarily mass-selected from photometric samples with confirmed spectroscopic redshifts, and cover fairly homogeneously the stellar mass and SFR ranges of the massive star-forming galaxy population at $0.5 < z < 2.7$. Over this redshift range, there is no particular bias at stellar masses $\log(M_\star/M_\odot) \gtrsim 10$, except for two factors. The first is a reduced Hα detection rate at the reddest colors (reflecting high dust obscuration or very low SFRs) and galaxies that are well below the "main sequence" (MS) of SFR at offsets of $\delta MS = \log(SFR/SFR_{MS}(M_\star, z)) \lesssim -0.85$ dex (Wisnioski et al. 2019). The second being a small excess of above-MS galaxies among the PHIBSS $z > 1$ CO samples. At lower stellar masses, the IFU and mm interferometry samples contain increasingly fewer galaxies with increasing redshift mainly because of (i) the additional $K$-band magnitude cut applied for the Hα surveys, and (ii) the lower galactic metallicities suppressing CO emission.

*Sample selection.* Building on RC41, we followed a similar strategy as detailed by G20 in expanding to the RC100 set. We start from the same subset of rotation-dominated ($v_{rot}/\sigma_0 > 2.3$) SFGs at $\log(M_\star/M_\odot) > 9.5$ and with $-0.6 < \delta MS < 1$, representing 65% of the full Hα+CO parent surveys (where the kinematic criterion makes the largest cut, ~25%). We removed galaxies for which residuals from night sky line emission in the near-IR affects part of their Hα line emission profile. We then excluded galaxies that are insufficiently spatially resolved in the data sets, and for which the radial coverage of detected line emission is too limited for our purposes. In applying the latter steps, we were less strict than G20. We combined seeing-limited + AO data sets for several more Hα targets, and included deeper observations of CO targets, which allowed the sample expansion to RC100. Given the often clumpy or irregular distribution in Hα and CO emission, the range of angular FWHM resolution of the observations (from ~0.2″ for SINFONI AO to 0.5"- 0.8″ for the other Hα and CO data sets), and the range of galaxy sizes, it is difficult to adhere to a specific set of S/N, size, and size/beam criteria.

Figure 1 shows the distribution of the RC100 sample in key galaxy properties, and compares it to the population of galaxies in the same redshift interval with stellar masses $\log(M_\star/M_\odot) \gtrsim 9.5$. For comparison, we show the sample of 240 of SFGs at similar redshifts from the KMOS$^{3D}$ survey (Wuyts et al. 2016, hereafter W16; green dots), as well as 3D-HST massive star-forming galaxies at similar redshifts (Skelton et al. 2014; Momcheva et al. 2016; grey dots). We split RC100 into a low redshift bin ($z = 0.6 - 1.2$, 33 galaxies) and a high redshift bin ($z = 1.2 - 2.5$, 67 galaxies). These ranges span approximately equal cosmic time intervals of 7.7-5 and 5-2.5 Gyr after the Big Bang. Compared to RC41, RC100 roughly doubles the number of objects at $\log(M_\star/M_\odot) \gtrsim 10.3$ and significantly extends the coverage to smaller sizes and higher central baryonic densities, appreciably reducing the bias towards large galaxies. The median effective radius is $<R_e> = 5.5$ kpc and the median baryonic surface density is $<\Sigma_{baryon}> = 10^{8.7}$ M$_\odot$kpc$^{-2}$. RC100 galaxies probe well the MS SFG population above $\log(M_\star/M_\odot) \gtrsim 10.3$ with a median offset ($\delta MS = SFR/SFR_{MS}$) of 0.26 dex (0.3 for RC41); the lower mass objects tend to lie above the MS. The RCs and dispersion profiles extend on average to $2.2R_e$ ($\pm 0.86R_e$, median $2R_e$), ranging from $1.3R_e$ to $4R_e$. This is slightly lower on average compared to RC41, and is a consequence of our less strict selection allowing objects with lower S/N (see above). The median beam FWHM for RC100 is 0.53", corresponding to a median physical scale of $0.75R_e$ ($\pm 0.38R_e$) at the corresponding redshifts.

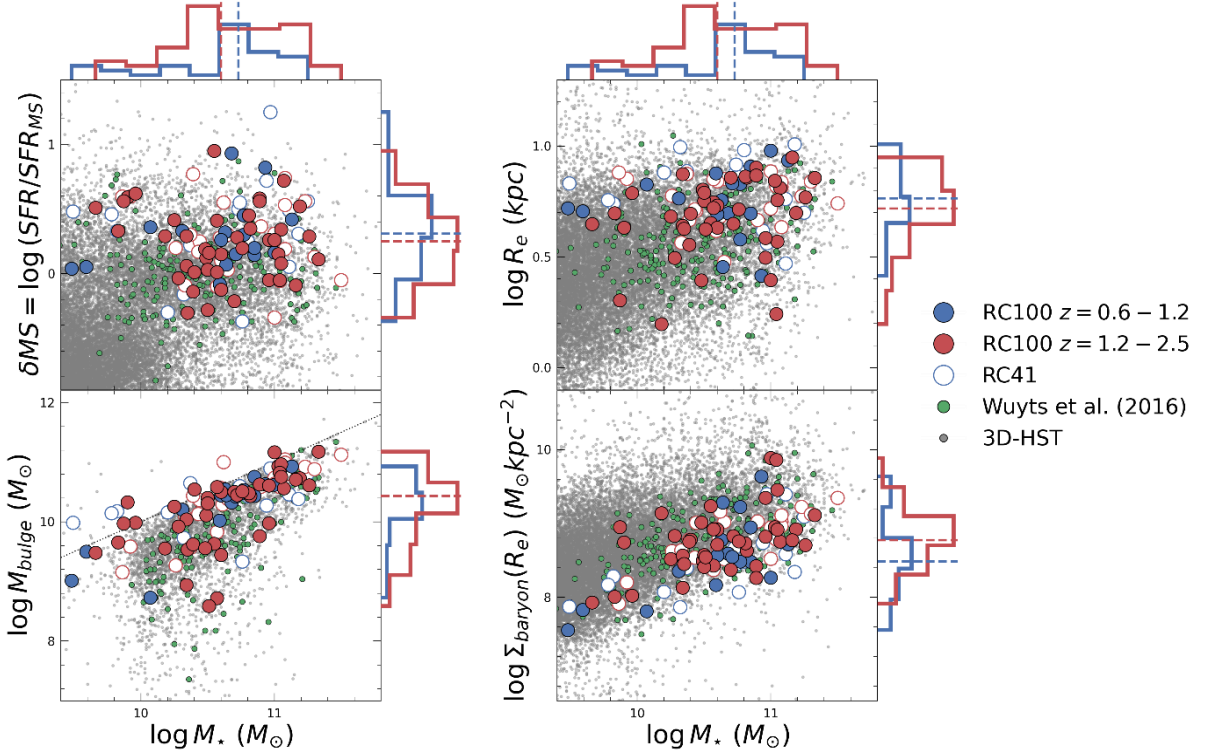

*Figure 1: locations of RC100 on the stellar-mass planes: offset from the SFR-MS (top-left), effective radius (top right), bulge mass (bottom-left,) and baryon surface density (bottom-right). RC100 galaxies are shown in blue (z = 0.6 − 1.2) and red (z = 1.2 − 2.5) circles, with galaxies included in RC41 (G20) as open circles. Histogram above and to the right show the frequency of RC100 galaxies for the two redshift intervals, with the dashed lines showing the median values. Green points show the 240 massive SFGs from the $KMOS^{3D}$ survey at z = 0.6 − 2.5 (W16). Background grey points are taken from the 3D-HST catalog for the same redshift range. RC100 values shown from kinematics, other values are based on photometry.*

*Major-axis rotation curves.* The dynamical mass of a rotationally supported, isolated disk galaxy can be best inferred from the projected rotation curve along its major axis, after correction for instrumental resolution ("beam smearing") and inclination. Circular motions are best captured along the major axis, where the effects of radial inflows/outflows and higher-order terms in the gravitational potential have lower impact (van der Kruit & Allen 1978; van der Kruit & Freeman 2011). Such effects would be more significant along the minor axis, where the line of sight circular motions tend to zero. In addition, the minor-axis is not detected as far out as the major-axis (in angular size), meaning it is more sensitive to the size of the PSF. Price et al. (2021) addressed these issues directly for RC41, modeling 2D velocity and dispersion maps and comparing it against 1D major axis RCs fitting. They showed that there is no significant difference when using 1D RC or 2D velocity maps, and that the latter is preferable for exploring the non-circular motions. Nonetheless, 2D maps are used to better infer the kinematic major axis and to maximize S/N out to large radii by using pseudo-slits depending on isovelocity contours. We construct the RCs following the most common approach extracting velocity centroids and velocity dispersion (from Gaussian fitting) for suitable slit cuts along the dynamical major-axis of the galaxy (see section 2.1 of G20 for further details). The RCs of RC100 exhibit a wide variety of rising, flat and falling shapes. Half of all RCs have flat slopes in the outer regions, and roughly a quarter show noticeable drops (≥ 10%) from the peak velocities. There are no major mergers in RC100, and at most 3 minor mergers. Low mass companions (of 0.05 − 0.1 × the mass of the central galaxy) are present in 8 of the cases, although actual evidence for interaction from spectra is not available for most of these cases (G20). As discussed in G20 such interactions could affect the characteristics and interpretation of the RC, but the clear point-symmetry of most of the RC100 speaks against such concerns (See G20, Section 3.1).

## 3. Methods

*Mass modelling.* We fit each individual galaxy in the RC100 with a mass model consisting of a turbulent thick disk with a constant intrinsic velocity dispersion, a stellar bulge, and a dark matter halo. Each of the mass components is described with a set of parameters, from which we forward-model the observed rotation curve (including beam-smearing and geometrical factors) and compare to the observational data. The mass model is described briefly in this Section, and a more detailed discussion can be found in Appendix A. However, the choice of fitting procedure can introduce biases and degeneracies inherent to the method. The inferred best-fit values can be susceptible to covariances between model parameters and/or prior constraints. Price et al. (2021) discusses this in detail in their Appendix A4, illustrating the degeneracies when using Bayesian analysis, in particular regarding $f_{DM}(R_e)$ and $M_{vir}$. To address this, we fit the velocity and dispersion profiles of each galaxy in RC100 with three fitting procedures including least-squares and Bayesian approaches. Each of these procedures have different covariances, and by comparing the three sets of best-fit values we can minimize their effect. The three methods used are described in this Section, and we find all three procedures to be in good agreement with one another, with no systematic bias.

The disks and bulges are modeled as de-projected Sérsic profiles with a total mass $M_{baryon}$ and bulge-to-total ratio $B/T = M_{bulge}/M_{baryon}$ (with circular velocities following Noordermeer 2008). The bulges are all spherical with Sérsic index $n_{s,bulge} = 4$ and projected effective radius $R_{e,bulge} = 1$ kpc. The disks are modeled as flattened spheroids described by an intrinsic axis ratio $q_{0,disk}$, Sérsic index $n_{s,disk}$ (hereafter $n_s$), and an effective radius $R_{e,disk}$ (hereafter $R_e$). Observationally, we assume the light traces the disk with a mid-plane distribution following the same profile as the mass distribution ($n_s$, $R_e$) with a constant mass-to-light ratio, and assume the bulge emits no light in the kinematic emission line tracers. In our default models we adopt spherically symmetric NFW dark matter density profiles (Navarro, Frenk & White 1996), with total mass $M_{vir}$ and concentration parameter $c = r_{vir}/r_s$. We account for adiabatic contraction of the halos only when stated so (Mo, Mao & White 1998). Additionally, we include pressure support correction using the self-gravitating disk formulation by Burkert et al. (2010) (see equation [8] in Appendix A). Finally, we consider galaxy-scale dark matter fractions, defined as the ratio of the DM and circular velocities at $R_e$:

$$[1] \quad f_{DM}(R_e) = \frac{V_{DM}^2(R_e)}{V_{circ}^2(R_e)}$$

which will also refer to as $f_{DM}$ for shorthand.

*Resolution corrections.* The intrinsic model rotational velocities and velocity dispersion are used to construct the beam-smeared, projected 1D rotation and dispersion curves along the major-axis, which we fit against the observational data. In all three methods we use a Gaussian resolution element (a spatial Point-Spread Function and spectral Line-Spread Function), but Methods (A) and (B) convolve the PSF on a 3D cube ($x_{sky}, y_{sky}, V_{los}$) before extracting velocities and dispersion from Gaussian fits, whereas Method (C) performs the convolution on the rotation velocities on the galactic plane (see Sections 3.1-3.3). In addition, Method (A) uses a Marquardt-Levenberg least-squares fitting and Methods (B) and (C) use Monte-Carlo Markov-Chains (MCMC) using the python package *emcee* (Foreman-Mackey et al. 2013) to explore simultaneously the posterior distributions of the parameters. We perform these three independent analyses using similar underlying mass models and priors, so that any deviations arise from the specific fitting procedure used and resolution correction. We find that the best-fit values are in broad agreement with one another considering the uncertainties.

*Priors.* We use both ancillary and simulation-based data to constrain our model parameters. Multiband photometry covering rest-UV to optical wavelengths and spectral energy distribution (SED) fitting following the procedures described by Wuyts et al. (2011) provide estimates for $M_\star$ (see also Förster Schreiber et al. 2009, 2018; Wisnioski et al. 2019). The scaling relations of Tacconi et al. (2020) provide the cold gas mass estimate, $M_{gas}$, given the redshift and stellar mass. The sum of these stellar and gas masses provides initial estimates for the total baryonic mass, $M_{baryon,init} = M_\star + M_{gas}$. The disk axis ratio $q_{0,disk}$ is taken to be $\sim 0.2 - 0.25$, a typical value for disks at these redshifts (van der Wel et al. 2014). The inclination, $R_e$ and Sérsic index of the disk are provided from 2D optical high-resolution HST imaging accounting for the finite intrinsic thickness of the disk, $q_{0,disk}$ (Lang et al. 2014 Tacchella et al. 2015). The concentration of the halo is determined from a scaling relation motivated from simulations (e.g., Dutton & Macciò 2014) with a dependence on redshift only (the assumed concentration has little impact on the best-fit values, see G20 Appendix A.4 for further discussion). For further discussion on the mass modelling and fitting procedures, we refer the reader to Appendices A, A.1 & A.2.

## 3.1 Marquardt–Levenberg (ML) gradient least-squares using DYSMAL (A)

For the first method considered here, we use a least squares approach to fit both the 1D major-axis rotation curve and dispersion curve via full forward modeling using our assumed mass model, following the methodology introduced in G20. The mass model is used to construct a 4D hypercube (over $x_{sky}, y_{sky}, z_{sky}, V_{los}$), which is collapsed along the line of sight according to the galaxy's inclination and convolved with both the spatial PSF and the spectral line spread function (LSF) to yield a model 3D cube including all observational effects. We extract 1D rotation and dispersion curves along the major-axis with Gaussian fitting using aperture extraction with the same slit/aperture positions and sizes.

We used ML gradient least-squares fitting for 4 free parameters, rising to 6 free parameters in rare cases, with the primary four being the total baryonic mass ($\log M_{bayron}$), bulge-to-total ratio ($B/T$), velocity dispersion ($\sigma_0$) and dark matter halo mass ($M_{vir}$). In some cases, the disk effective radius ($R_e$) and inclination ($i$) were also used as free parameters, constrained by ancillary imaging data. In most cases, the inclination and effective radius are fixed from photometric-determined values. Qualitative constraints were recovered from the 1D $H\alpha$ or $CO$ flux observations for the Sérsic index of the disk ($n_s$), but they were not explicitly fitted. Adiabatic contraction of the halo was also considered for all cases, and the dark matter fractions were calculated as an average of a regular NFW halo and a contracted NFW halo.

## 3.2 MCMC using DysmalPy (B)

We perform an MCMC analysis using the *emcee* python package (Foreman-Mackey et al., 2013) on the 1D velocity and dispersion profiles following the procedures used in Price et al. (2021). We extract these curves using the same full forward-model approach used in Method A, using the same instrumental and spectral parameters. The key points of this fitting are summarized here, and we refer the reader to Price et al. (2021) for a complete description (their Section 3.2). The MCMC parameter space exploration was conducted on 4 free parameters: $\log M_{bayron}$, $R_e$, $\sigma_0$, $f_{DM}(R_e)$, with Gaussian priors for $\log M_{baryon}$ centered on $M_{baryon,init}$ derived from ancillary data, and $R_e$ centered on inferred values from method A (either from HST imagery or least-squares fitting). Flat priors are used for $f_{DM}(R_e)$ and $\sigma_0$ (see Table 2 of Price et al. 2021). The $B/T$ ratio is determined from Method A and taken to be constant. The best-fit values for this analysis are taken to be the maximum a posteriori (MAP) values of the fit parameters, based on a joint analysis of the posteriors for all 4 free parameters using a multi-dimensional Gaussian kernel density estimator to account for degeneracies between the parameters. The $1\sigma$ uncertainties are determined from the shortest 68% intervals of the individual parameters' marginalized posterior distributions.

## 3.3 MCMC using galaxy-plane beam-projection (C)

For the third method we used a similar Bayesian fitting approach as Method B, using the python package *emcee* (Foreman-Mackey et al., 2013), but with an alternate treatment of the beam-smearing. The 1D major-axis beam-smeared RC is fitted to the observational data from which the best-fit parameters are inferred. For a given set of model parameters, the intrinsic circular velocity is calculated in the galactic plane. A circular beam with a Gaussian PSF intersects the plane with an effective projected area depending on the inclination of the galaxy on the sky, increasing for high inclinations ($i = 0$ face-on). The beam-smeared velocity is given by a flux-weighted average of the line-of-sight circular velocity over the projected area of the beam, and the dispersion as its standard deviation. We extract points along the major axis and compare them to the observed data points. The fitting procedure performs a least-squares minimization for the rotation and dispersion curve simultaneously.

We used a MCMC parameter exploration with 4 free parameters: $\log M_{baryon}$, $R_e$, $\sigma_0$ and $\log M_{vir}$. The dark matter fractions are calculated at each step of the fitting process from the given set of parameters (following eq. [1]). We use Gaussian priors for $\log M_{baryon}$ centered on $M_{baryon,init}$ derived from ancillary data, and for $R_e$ centered on our results from Method A. Wide Gaussian priors are used for $\log M_{vir}$ centered on abundance-matching relations (Moster et al. 2018), and a flat prior is used for $\sigma_0$. The best-fit values from these fits are taken as the median values, with $1\sigma$ uncertainties derived from the closest 68% intervals of the parameters' posterior distributions.

## 4. Results

***Comparison of the three analysis methods.*** We derive the best fitting kinematic parameters, $f_{DM}(R_e)$, $M_{baryon}$, $R_e$, $B/T$ and $\sigma_0$, for each of the RC100 galaxies using all three methods. The inferred values are found to be in good agreement within the uncertainties, and we take the average value of all three methods to be used in the following analysis (See Appendix B for the complete table). Comparing all sets of results, we calculate the absolute difference between pairs of values in terms of the combined uncertainty: $\Delta X_{ij} =$

$|X_i - X_j|/\sqrt{\Delta X_i^2 + \Delta X_j^2}$. Here, $i, j$ are indices for the three fitting Methods (A, B and C), $X_i$ is a model parameter for a galaxy for the method $i$, and $\Delta X_i$ is the uncertainty in the parameter. A value of $\Delta X_{ij} < 1$ would be interpreted as a good agreement. On average, all model parameters reach this criterion. Furthermore, the Pearson coefficients between all sets of results are $\geq 0.72$ up to $0.94$. For a more detailed discussion on the comparisons see Appendix C. Finally, for the lensed galaxy *J0901+1814* we adopt the inferred values from a detailed kinematic modeling of the de-lensed rotation curve reaching spatial a resolution of 600 pc in the source plane (Liu et al., in prep.)

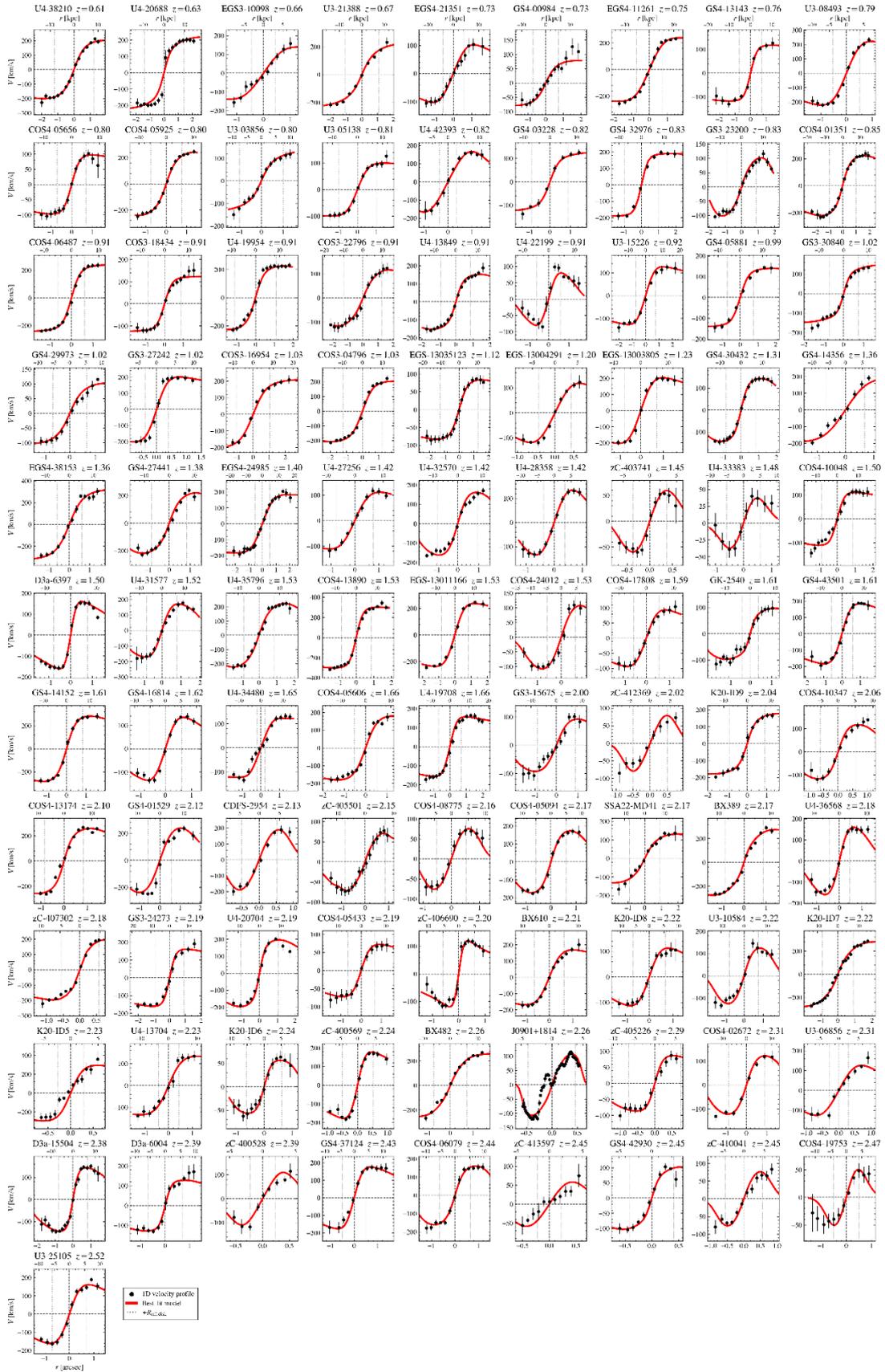

Figure 2: Rotation curves and best-fit model results (A, red curve) for the RC100 sample, ordered by redshift from top-left to bottom-right. In each panel, the bottom axis is in units of arcsec and the top in kpc.

## 4.1 Dependance of $f_{DM}(R_e)$ on redshift

*Low galaxy-scale DM fractions at high-z.* While in the Local Universe disk galaxies are mostly DM dominated, at higher redshifts the instance of baryon dominated disk galaxies greatly increases. Looking at the circular velocities at $R_e$, $V_c(R_e)$, compared to $f_{DM}(R_e)$, $z\sim 0$ disks show an anti-correlation with DM fractions, where only the more massive galaxies ($V_c \gtrsim 250$ km/s) become baryon dominated (Barnabé et al. 2012; Dutton et al. 2013; Martinsson et al. 2013, Courteau et al. 2015; Cecil et al. 2016). As demonstrated by G20, and reaffirmed by our increased statistics, RC100 galaxies become increasingly baryon dominated at higher redshifts, even at much lower circular velocities ($V_c \approx 100$ km/s).

Figure 3 shows $f_{DM}(R_e)$ vs. the circular velocity at $R_e$ (corrected for pressure support) for the two redshift bins: $z = 0.6 - 1.2$ (blue), $z = 1.2 - 2.5$ (red). The median DM fractions of $z = 0.6 - 1.2$ RC100 galaxies is $<f_{DM}> = 0.38 \pm 0.23$ while for $z = 1.2 - 2.5$ the median DM fractions drops to $0.27 \pm 0.18$, similar to that of "maximal disks" ($f_{DM}(R_e) \leq 28\%$). At $z \sim 1$, the share of RC100 galaxies with fractions lower than "maximal disks" is only roughly 33%. Additionally, the scatter in $f_{DM}(R_e)$ for a given circular velocity is greater at $z \sim 2$ than at $z \sim 1$, across a wide range of circular velocities. This can be seen by the corresponding Spearman Rank coefficients, which are -0.35 and -0.06 at $z \sim 1$ and $z \sim 2$ respectively, pointing towards a large galaxy-to-galaxy variance when determining their DM content, where some accumulate DM efficiently while others do not. In fact, at $z \sim 2$ the dark matter content of disk galaxies is much closer to that of local early-type galaxies (ETGs, green triangle; Cappellari et al. 2011, 2013), their possible descendants. These baryon-dominated SFGs at $z \sim 2$ could have depleted their gas reservoirs rapidly and quenched their star formation, evolving into local ETGs, while the $z \sim 2$ DM dominated SFGs managed to retain their star formation processes and settle as LTGs. Finally, our sample is augmented at $z \sim 1$ with nine low-mass, low central baryon surface density SFGs observed with MUSE (Bouché et al. 2021), continuing the trend set by $z \sim 0$ disks.

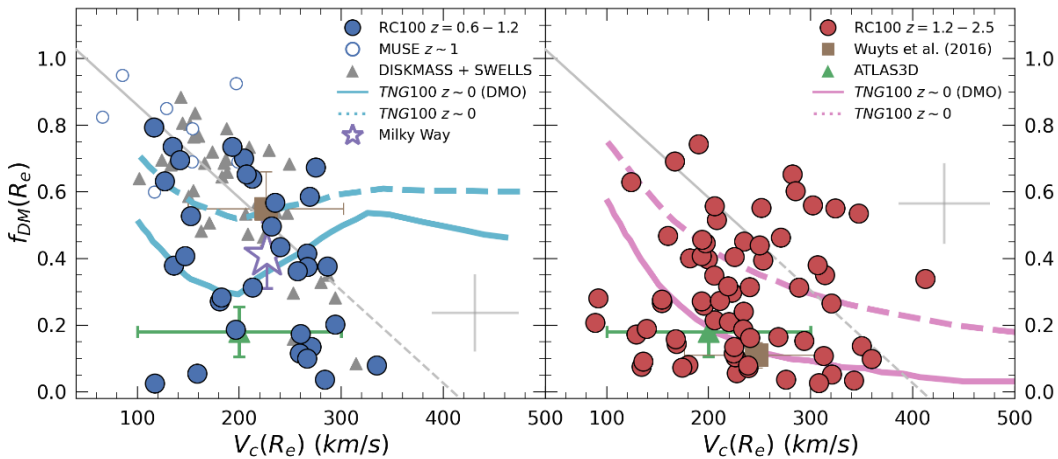

*Figure 3: Galaxy scale dark matter fractions of RC100 galaxies at $z = 0.6 - 1.2$ (left panel, blue circles) and $z = 1.2 - 2.5$ (right panel, red circles), with $V_c(R_e)$ the circular velocity at $R_e$. The average uncertainties are given in large grey crosses. RC100 is extended at $z \sim 1$ with nine massive SFGs at observed with MUSE towards lower surface density SFGs (open blue circles, Bouché et al. 2021). $z = 0$ disk galaxies from the SWELLS and DiskMass surveys are shown in grey triangles (Barnabé et al. 2012; Martinsson et al. 2013), with a linear trend in a solid-dashed grey line, and the Milky Way is shown for comparison (purple star; Bovy & Rix 2013; Bland-Hawthorn & Gerhard 2016). In both panels the brown solid box shows the average values from KMOS$^{3D}$ for the*

*same redshift bins (W16) and the green triangles shows average values for 260 $z = 0$ ETGs from ATLAS-3D (Cappellari et al. 2011, 2013). TNG100 values for the hydrodynamical (DM-only) version is given by dashed (solid) line at $z = 0$ and $z = 2$.*

*Comparisons to simulations.* Cosmological simulations provide predictions for the dark matter fractions, which are found to be in tension with the RC100 results $f_{DM}(R_e)$ at higher redshifts. The *Illustris TNG100* cosmological simulation calculates $f_{DM}$ for massive disk galaxies in two different ways, to estimate the effect of the baryons on the dark matter: firstly with a full hydrodynamical simulation including the reaction of DM to the baryons (dashed lines in Figure 3), and secondly by embedding a formed galaxy in a halo resulting from a DM-only version of the simulation (Solid lines in Figure 3; Lovell et al., 2018). Generally, the full hydrodynamical simulations recover higher $f_{DM}$ values than the DM-only versions, due to the contraction of DM. These predictions match well for the Local Universe, but at $z \sim 2$ the hydrodynamical predictions of *TNG100* do not reach the very low values of $f_{DM}(R_e) < 0.2$ (Despite the simulated galaxies having typically smaller $R_e$, see Übler et al. 2021). ***These lower observed dark matter fractions compared to the TNG100 values could point toward stronger, more effective, feedback than adopted in the simulations.***

Stellar and AGN feedback, with efficient heating from dynamical friction on merging satellites, can cause the DM to form cores in initially cuspy profiles (De Souza et al. 2011; Ogiya and Mori 2011; Governato et al. 2012; Lazar et al. 2018; Schulze et al. 2020; Dekel at al. 2021; Li et al. 2022; Ogiya and Nagai 2022). The *MAGNETICUM* simulation, which includes higher resolution and more detailed feedback processes, have shown that dark matter fractions fall from an average of $< f_{DM} >_{z=0}= 0.36 \pm 0.10$ to $< f_{DM} >_{z=2}= 0.10 \pm 0.05$ (Remus et al. 2017), in better agreement with our observations. In the *FIRE* simulation, detailed stellar feedback processes have formed cores in the DM halo of massive dwarf galaxies at $z = 0$ (Chan et al. 2015; Lazar et al. 2018), but more massive galaxies (akin to RC100) remain cuspy. The dominant mechanism for massive galaxies is more likely to be AGN feedback rather than stellar-driven feedback.

*Redshift dependence.* Based on these findings, our first major conclusion is that dark matter fractions decrease as the redshift increases, more strongly than predicted from cosmological simulations. The median dark matter fractions decline from $< f_{DM}(R_e) > = 0.38$ at $z = 0.85$ to $< f_{DM}(R_e) > = 0.17$ at $z = 2.44$, showing a clear declining trend. Our sample is extended at $z = 0$ by disk galaxies from the DISKMASS and SWELLS surveys. *TNG100* (full hydrodynamical) predictions for $M_\star = 10^{10}$ M$_\odot$ (solid) and $M_\star = 10^{11}$ M$_\odot$ (dashed) galaxies overestimates DM fractions by a factor of ~2, mostly at the higher redshifts. As the majority of RC100 galaxies are have stellar masses between these values, they should populate the space between the *TNG100* curves, yet they are consistently below them at all redshifts. We can describe the trend set by the RC100 median DM fractions reasonably well using a power law of the form $< f_{DM}(R_e) > (z) = a (1 + z)^{-b}$, where we find $a = 0.75 \pm 0.23$ and $b = 1.07 \pm 0.33$. This relation also matches well with the $z = 0$ disks, even though we do not explicitly fit for them. Since RC100 does not necessarily represent the entire SFG population at high-z, this relation should be taken heuristically.

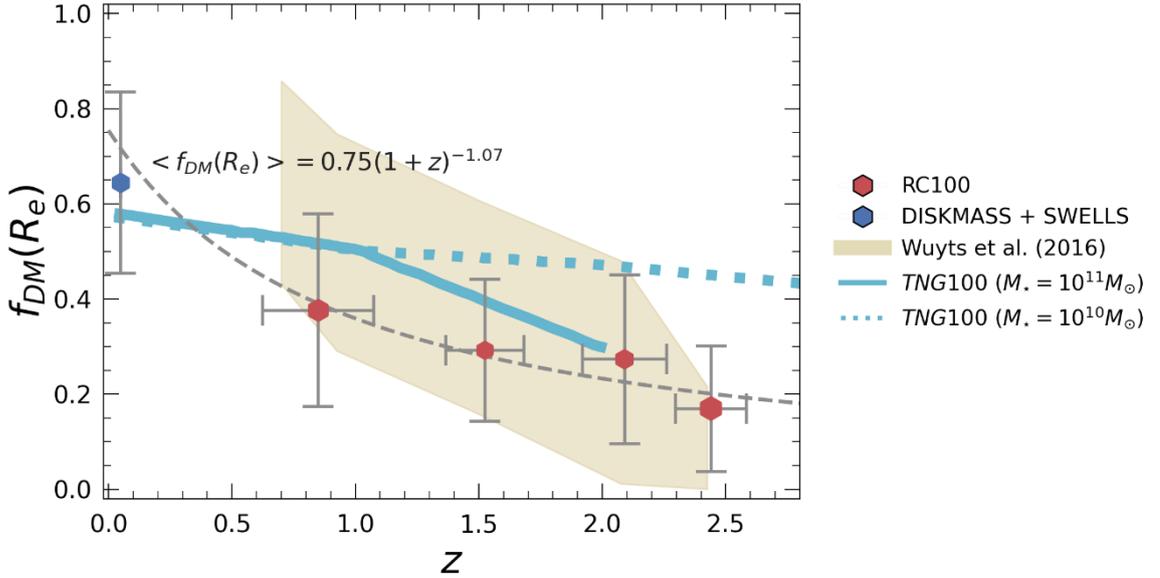

*Figure 4: Dark matter fractions of massive SFGs decrease with redshift, from 0.7 at $z = 0$ to 0.37 at $z = 0.85$ and 0.17 at $z = 2.44$. Hexagons show median values of $f_{DM}(R_e)$ binned at finer redshift bins, with the error bars showing the standard deviation and hexagon size representing uncertainty of the mean for each bin ($\sigma/(N-1)$). Red hexagons show RC100 galaxies and the blue hexagon shows $z = 0$ disks from the SWELLS and DiskMass surveys. The dashed grey line is a best-fit power law for the RC100 alone, with the values given in the plot. The shaded region shows the values from W16 for 240 SFGs at similar redshift ranges, which are higher at $z \sim 1 - 1.5$ due to the typical larger disk sizes of RC100. TNG100 DM fractions from the full hydrodynamical suite are shown by blue lines for disk galaxies with $M_\star = 10^{10} M_\odot$ (dashed) and $M_\star = 10^{11} M_\odot$ (solid).*

## 4.2 Baryonic surface density

***Strong anti-correlation with baryonic surface density.*** One of the key results of G20 and W16 is the tight correlation between $f_{DM}(R_e)$ and $\log \Sigma_{baryon}(R_e)$, which is also present in our extended RC100 data set (See Figure 5). We find that the anti-correlation is observed in both redshift bins of RC100, although it is stronger among the lower redshift galaxies (Spearman coefficient of -0.84) than at higher redshifts (Spearman coefficient = -0.64). This is further caused by the large scatter noticed in Figure 3, pointing towards larger galaxy-to-galaxy variance at higher redshifts. RC100 galaxies span almost 2 orders of magnitude in surface density, with the higher redshift galaxies having larger $\Sigma_{baryon}(R_e)$ on average. Continuing our analysis in Section 4.1, roughly 67% of $z \sim 1$ RC100 galaxies occupy a similar parameter space as local spirals, while only 33% overlap with the distribution of $z = 0$ ETGs (green crosses). At high-z, it is reversed: roughly two thirds of the galaxies occupy the same space as $z = 0$ ETGs (green triangles), with some even nearing ETGs at $z \sim 1.7$ (VIRIAL, Mendel et al. 2020). ***High-z SFGs are more similar to local early-type galaxies in their central DM content and baryonic surface densities than to local late-types.*** This leads to question whether such highly baryon-dominated SFGs quench their star formation and become ETGs at later stages of their evolution. In both redshift intervals, *TNG100* (full hydrodynamical) overestimates the DM fractions especially at high baryon surface density, deviating from the W16 relation above $\log \Sigma_{baryon} \gtrsim 10^{8.5} M_\odot \text{kpc}^{-2}$. The *MAGNETICUM* simulation reaches lower $f_{DM}$ at $z \sim 2$, in better agreement with our observations, but is also overestimating $f_{DM}(R_e)$ at high baryon surface density. These findings further suggest that some mechanism allows baryons to accumulate efficiently in the central regions of the galaxy (reaching high surface density), either without contracting the surrounding dark matter at all or by effectively redistributing the central dark matter, so the central regions become baryon dominated.

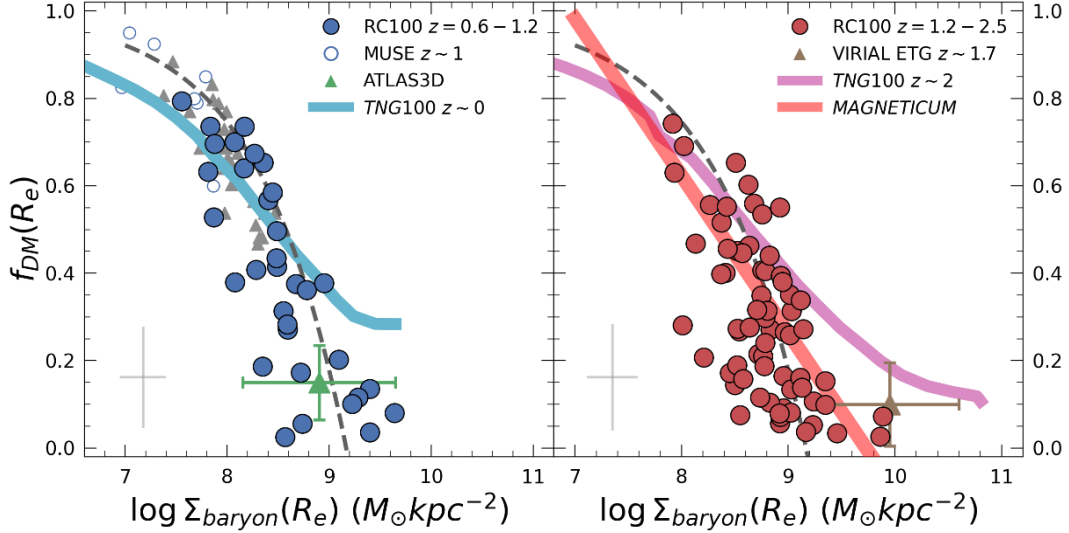

*Figure 5: Anti-correlation of Dark matter fractions with baryon surface density. RC100 galaxies at $z = 0.6 - 1.2$ (left panel, blue circles) and $z = 1.2 - 2.5$ (right panel, red circles) have Spearman coefficients of (-0.84) and (-0.64), respectively, indicating a strong anti-correlation. The median uncertainties of RC100 are shown in grey crosses. RC100 galaxies lie on a similar trend found by W16 from modeling the inner RCs of 240 massive star-forming galaxies (dashed curve), also occupied by $z \sim 1$ nine MUSE SFGs (open blue circles), as well as local $z = 0$ LTGs (grey triangles) and $z = 0$ ETGs (green triangle). $z \sim 1.7$ massive quiescent galaxies from the VIRIAL survey (25 galaxies, brown triangle; Mendel et al. 2020) have higher baryon surface density and do not overlap with (almost) any RC100 galaxy. Blue and pink lines show TNG100 values for $z = 0$ and $z = 2$, respectively (Lovell et al. 2018).*

### 3.24.3 DM deficit compared to Stellar-Mass Halo-Mass relations

*A deficit in dark matter content.* We investigate the dark matter content on galactic scales as inferred from our dynamical fitting, compared to the amount of DM arising from stellar-mass-halo-mass (SMHM) scaling relation for NFW halos (Moster et al. 2018). As the 3D enclosed mass for an NFW profile is increasing with disk size ($M_{NFW} \propto r^2$ at small radii), larger disks that probe further into the halo contain more dark matter mass. To counteract this effect, we consider the projected DM mass surface density on the area defined by $R_e$:

$$[2]\ \Sigma_{DM}(R_e) = \frac{M_{DM}(R_e)}{\pi R_e^2}$$

For an NFW halo the mass within $R_e$ is a simple function of $(c, M_{vir}, R_e)$ that can be calculated directly. By comparing the two we can estimate the amount of missing DM content. We therefore calculate this property twice: once using the kinematically inferred $M_{vir}$ from our fitting, and once by using $M_{vir}(M_\star, z)$ from the SMHM scaling relations of Moster et al. 2018. We use stellar masses derived from 3D-HST photometry for each of our RC100 galaxies and correct it for the kinematically inferred baryon mass, $M_{\star,fit} = \frac{M_{baryon,fit}}{M_{\star,HST}+M_{gas}} M_{\star,HST}$, with $M_{gas}$ the gas mass determined from scaling relations (Tacconi et al. 2020). The kinematically corrected stellar masses are used to match the total halo mass for the galaxy, $M_{vir}^{SMHM}$, which is then used to get $\Sigma_{DM}^{SMHM}(R_e)$.

Massive SFGs with low $f_{DM}(R_e)$ can have up to an order of magnitude lower $\Sigma_{DM}(R_e)$ than predicted from SMHM relations, and this deficit is more evident for galaxies with massive bulges. Halos following the SMHM relation will generally have a fairly constant $\Sigma_{DM}(R_e)$, of the order of $\sim 10^8 - 10^{8.5} M_\odot kpc^{-2}$ (shown as a grey shaded area in Figure 6), and at low DM fractions we know that $\Sigma_{DM}(R_e) \propto f_{DM}(R_e)$. Roughly 25% of all RC100 galaxies have dark matter surface densities that are at least 0.5 dex lower than predicted, the majority of them at $z > 1.2$. We define the offset in DM surface density as:

$$[3] \Delta \log \Sigma_{DM}(R_e) = \log \Sigma_{DM}(R_e) - \log \Sigma_{DM}^{SMHM}(R_e)$$

Negative values note a deficit in the DM content within $R_e$. The median offset for RC100 at $z \sim 1$ ($z \sim 2$) is $-0.06$ dex ($-0.18$ dex), meaning half of RC100 galaxies have removed more than 33% of their original DM mass within $R_e$ at $z \sim 2$. In terms of the baryon surface density (Figure 6, left panel), we see no correlation at either redshift interval (Spearman coefficients $\approx -0.02$). Even so, DM deficits seem to be more frequent above some baryon surface density threshold. For $\Sigma_{baryon}(R_e) > 10^{8.5} M_\odot kpc^{-2}$ the median offset is $< \Delta \log \Sigma_{DM}(R_e) (R_e) >= -0.22$ dex, while at lower baryon surface densities there is an abundance of DM, with a median offset of $+0.10$ dex. This is even more apparent when splitting also by redshift: at $z \sim 2$ disks with $\Sigma_{baryon}(R_e) > 10^{8.5} M_\odot kpc^{-2}$ have a median offset in DM surface density of $-0.27$ dex, while at lower baryon surface density it is only $< \Delta \Sigma_{DM}(R_e) >= -0.08$ dex. At $z \sim 1$ there is a similar behavior, with a median offset of $-0.11$ dex ($+0.18$ dex) for high (low) baryon surface density (see Table 2).

*Bulge mass.* The emergence of a deficit of DM can be seen more clearly with respect to the bulge mass. In Figure 6 (right panel), it is clear that the deviation from the expected SMHM (shaded region) exists mostly for bulge masses above $10^{10} M_\odot$. We find no correlation between $\Sigma_{DM}(R_e)$ and $M_{bulge}$, but we notice that having a massive bulge is linked with a deficit of DM, while having a smaller bulge with an abundance of DM. The median offset for $M_{bulge} > 10^{10} M_\odot$ is $-0.13$ dex while it is $+0.09$ dex for galaxies with less massive bulges. This effect is also greater at $z \sim 2$, where the median offset is $-0.30$ dex ($+0.06$ dex) for bulge masses above (below) $10^{10} M_\odot$, and much milder at $z \sim 1$ the offset is $-0.09$ dex ($+0.18$ dex) respectively (see Table 2). In a different way to find such threshold bulge mass, we slice the data set at different values of $M_{bulge}$ and look for the value at which the Spearman coefficient becomes statistically significant ($p < 0.05$). We find that this occurs at $M_{bulge,thresh} = 10^{10.5} M_\odot$, yielding a correlation coefficient of $-0.39$ corresponding to an anti-correlation. ***$z \sim 2$ SFGs with high baryon surface density and bulge mass are more likely to have a deficit of approximately 50% less DM compared to SMHM relations within $R_e$.***

However, determining bulge masses at these redshifts is uncertain at these redshifts, due to the low spatial resolution. Resolution effects can mask out some of the contribution of the bulge to the RC, which makes the high-resolution adaptive optics (AO) subset of RC100 to be a very valuable source for determining bulge masses. For these AO galaxies (shown by the stars in Figure 6), the median offset for galaxies with bulge masses above (below) $10^{10} M_\odot$ $-0.47$ $(+0.15)$ dex. The incidence of a detected broad component linked with an AGN increases dramatically at the same bulge mass cutoff ($\sim 30\% - 50\%$, Genzel et al. 2014; see also Förster Schreiber et al. 2019). Also, it is well known that bulge mass is correlated with the central BH mass (Madau & Dickinson, 2014). Rapid bulge growth at $z \sim 2$ can thus lead to two opposite effects: (1) contracting the surrounding DM towards the inner regions of the galaxy by deepening the gravitational well, and (2) increased feedback driving DM from the inner regions to the outer parts of the halo. The efficiency of either mechanism is key to determining the final state of the DM halo. Inflow of baryons during the build-up of the bulge can heat up the DM via dynamical friction, making AGN feedback more effective (El Zant et al. 2001; Dekel et al. 2021; Ogiya et al. 2022).

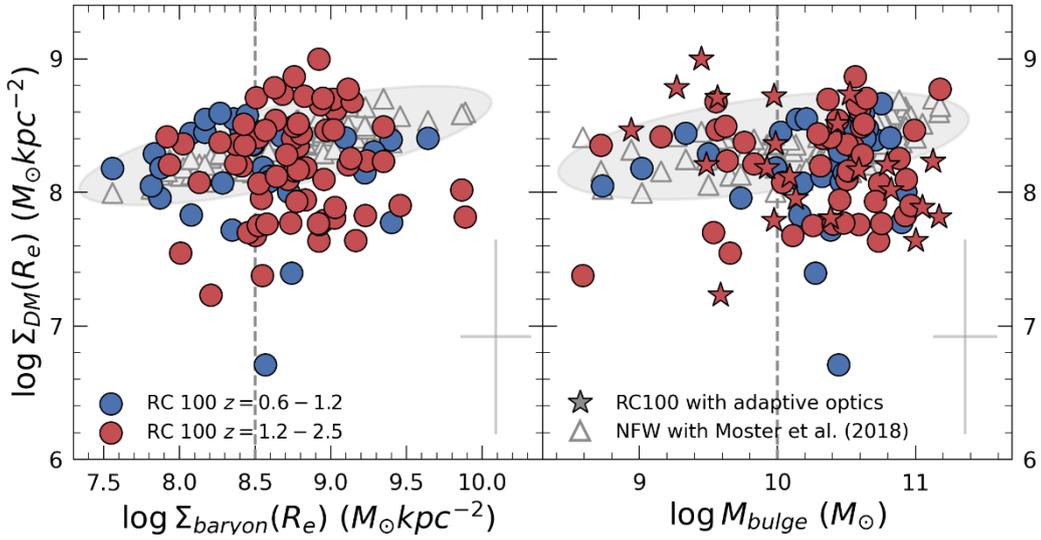

*Figure 6: Dark matter surface density as a function of baryonic surface density (left) and kinematically inferred bulge mass (right) for RC100 in the low (blue, $z = 0.6 - 1.2$) and high (red, $z = 1.2 - 2.5$) redshift intervals. RC100 galaxies with adaptive optics are shown as stars (right panel). Open grey triangles show the dark matter surface densities assuming the Moster et al. (2018) SMHM scaling relations and an NFW halo. At high $\Sigma_{baryon}(R_e)$ and bulge mass (given by the vertical dashed line) DM deficits are more abundant. Only a few galaxies lie significantly above the predicted values, but roughly 25% of the galaxies have a large DM deficit of up to an order of magnitude.*

| $\Delta \log \Sigma_{DM}(R_e)$ $= \log \Sigma_{DM}(R_e)$ $- \log \Sigma_{DM}^{SMHM}(R_e)$ | $\Sigma_{baryon}$ $/M_\odot kpc^{-2}$ | | $M_{bulge}/M_\odot$ (w/ AO) | | $\Sigma_{SFR}$ $/M_\odot\, kpc^{-2} yr^{-1}$ | |
|---|---|---|---|---|---|---|
| | $< 10^{8.5}$ | $\geq 10^{8.5}$ | $< 10^{10}$ | $\geq 10^{10}$ | $< 10^{0.25}$ | $\geq 10^{0.25}$ |
| $RC100$ | $-0.09$ | $+0.1$ | $-0.22$ | $-0.08$ | $-0.11$ | $-0.06$ | $-0.3$ |
| $z = 0.6 - 1.2$ | $-0.06$ | $+0.19$ | $-0.11$ | $-0.08$ | $+0.05$ | $-0.02$ | $-0.13$ |
| $z = 1.2 - 2.5$ | $-0.18$ | $-0.09$ | $-0.27$ | $-0.14$ $(+0.15)$ | $-0.3$ $(-0.47)$ | $-0.14$ | $-0.38$ |

Table 2: *Median values of the dark matter offset, $\Delta \log \Sigma_{DM}(R_e) = \log \Sigma_{DM}(R_e) - \log \Sigma_{DM}^{SMHM}(R_e)$, in different parameter ranges of RC100. In general, at $z \sim 2$ the offset is lower than in $z \sim 1$, indicating a larger deficit in DM within $R_e$ at higher redshifts. High baryon surface density, bulge mass, SFR surface density and hosting an AGN all have a larger median offset, when looking at the same redshift range. The values given in parentheses for $M_{bulge}$ address only the 22 galaxies including adaptive optics observations.*

### 4.4 SFR surface density

***SFR surface density linked with DM deficit.*** Strong feedback is needed to drive DM outwards from the center of the galaxy, which can be traced by the SFR. Bouché et al. (2021) showed that DM cores are linked with higher SFR surface density at $z \sim 1$ low surface density disk galaxies, and the *FIRE* and *MAGNETICUM* simulations shown stellar and AGN feedback can create low DM fractions. The SFRs of the RC100 galaxies range over two orders of magnitude, from $3 - 650\ M_\odot\ yr^{-1}$. We look at the average surface density of the SFR within $R_e$, $\Sigma_{SFR} = \frac{SFR}{\pi R_e^2}$, which spans almost three orders of magnitude from $0.02 - 18\ M_\odot\ yr^{-1}\ kpc^{-2}$ for all RC100 galaxies. A deficit in DM is found more frequently in galaxies with higher SFR surface density. This is seen in Figure 7 by the number of "red" points found below the grey shaded region representing the SMHM "expected" values. In the top two panels of Figure 7, we bin $\Sigma_{DM}(R_e)$ for several bins of $\Sigma_{baryon}(R_e)$ and $M_{bulge}$, and calculate the median $\Sigma_{SFR}$ at each bin. The drop from the SMHM predicted values is more apparent at higher SFR surface densities.

The median offset in DM surface density is $< \Delta \log \Sigma_{DM}(R_e) > = -0.29\ (-0.06)$ dex above (below) $\Sigma_{SFR} = 10^{0.25} M_\odot\ yr^{-1}\ kpc^{-2}$. At $z \sim 2$ this effect is even greater, with a median offset of $-0.38$ (-0.14), respectively (see Table 2). We find that a correlation becomes statistically significant for $\Sigma_{SFR} \geq 10^{0.4}\ M_\odot yr^{-1} kpc^{-2}$, yielding a Spearman correlation coefficient of -0.45. High SFR surface density correlates with a larger deficit in DM. Naturally, high $\Sigma_{SFR}$ leads to stronger stellar feedback. If the baryonic mass is small while the SFR is high, we would expect adiabatic contraction to be less efficient and the feedback leading to larger deficits in the DM. However, we see the opposite. This might be because stellar feedback is considered to be less effective in massive SFGs as it does not inject enough energy (Dekel & Silk 1986). A second explanation would incorporate the fact that SFR and bulge mass correlate with AGN strength, which is considered to be more efficient for such galaxies (Madau & Dickinson 2014; Förster Schreiber et al. 2019). Feedback from AGN can affect the form of the DM distribution (Waterval et al. 2022). When considering AGN feedback, larger baryonic masses would lead to stronger feedback and a larger deficit in DM, matching with our observed trends.

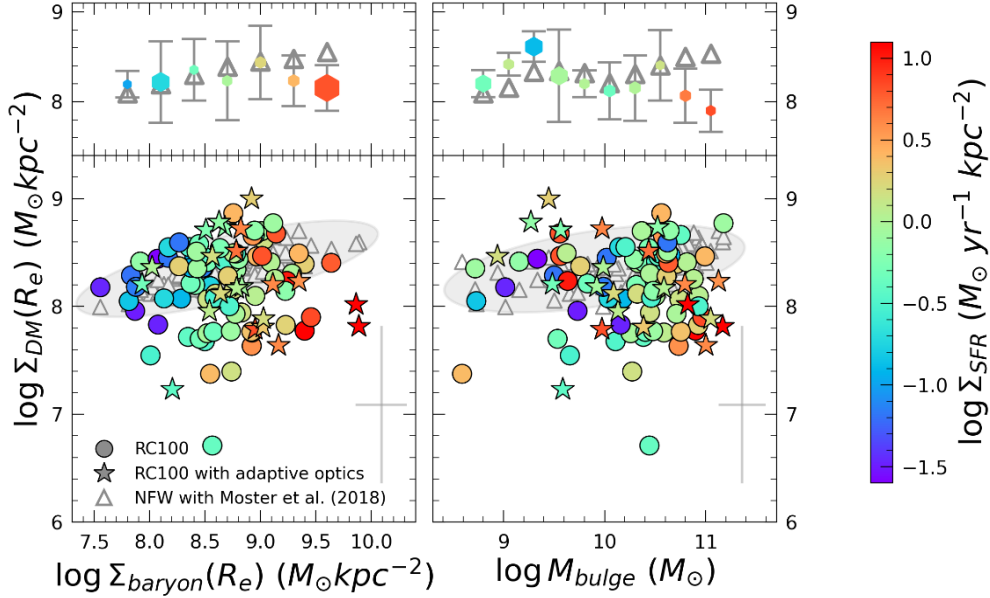

*Figure 7: Same as Figure 6, now color coded by SFR surface density for the range of RC100 galaxies. The top row shows binned median values with error bars showing the standard deviation and the hexagon size showing the uncertainty of the mean ($\sigma/\sqrt{N-1}$). The DM deficit at high baryonic surface density and bulge mass is linked with high $\Sigma_{SFR} \gtrsim 10^{0.25} M_\odot yr^{-1} kpc^{-2}$. The binned values show the drop more clearly for $\Sigma_{baryon}(R_e) \gtrsim 10^9 M_\odot kpc^{-2}$ and $M_{bulge} \gtrsim 10^{10} M_\odot$, both occurring at high $\Sigma_{SFR}$.*

## 4.5 Dark matter cores at $z \sim 2$

*Low DM fractions as cored distributions.* The DM deficits may reflect cored DM profiles rather than cuspy distributions, such as classical NFW profiles (c.f. G20). The inferred dark matter within $R_e$ is constrained by the observed kinematics, but the functional form of the dark matter halo, however, is a choice of the model. If within the disk effective radius there is less dark matter than expected, then assuming NFW and integrating out to the virial radius will give a total halo mass that can be much lower than predicted by the SMHM relation. For very low DM fractions this would imply that the DM is outweighed by baryons on the virial scale, with $M_{baryon}/M_{vir} \gtrsim 1$, which seems highly improbable in a standard $\Lambda CDM$ cosmology. Alternatively, we can assume a total halo mass given by the SMHM relation matching the kinematically inferred $M_\star$, and use a generalized version of the NFW profile with a varying inner slope:

$$[3] \quad \rho_{gNFW}(r) = \frac{\rho_o}{(r/r_s)^\alpha (1+r/r_s)^{3-\alpha}}$$

with $\rho_0$ the characteristic density, $r_s$ the scale radius of the halo and $\alpha$ is the inner slope. In this case, the total mass would be given by integrating the density profile all the way to the virial radius:

$$[4] \quad M_{vir}^{SMHM} = \int_0^{r_{vir}} 4\pi r^2 \rho_{gNFW}(r)$$

Taking the same concentration parameter used in this analysis with the halo mass from the SMHM relation, this equation can be solved numerically to derive $\alpha$. A cored density distribution would naturally explain the low central dark matter fraction (and dark matter deficit), while still maintaining the predicted total virial mass of the halo. For this purpose, we can look at the "effective slope", that is, the slope that would give the same amount of kinematically inferred dark matter within $R_e$ for a total mass determined from the SMHM relation. Figure 8 plots the dark matter fractions as a function of this "effective slope", showing that as the halo profile tends to a core ($\alpha \to 0$) the dark matter fractions approach $f_{DM}(R_e) <$ 0.2, and that profiles cuspier than NFW ($\alpha > 1$) result in high $f_{DM}(R_e)$. This is also redshift dependent, as 25% of RC100 galaxies at $z \sim 2$ have a slope of $\alpha < 0.5$ while this is true for only 10% of $z \sim 1$ galaxies. The median slope for $z \sim 1$ galaxies is 0.93, while for the $z \sim 2$ galaxies it is 0.77. This picture is consistent with our previous results of low DM fractions at higher redshifts, as cored DM distributions would yield lower $f_{DM}(R_e)$.

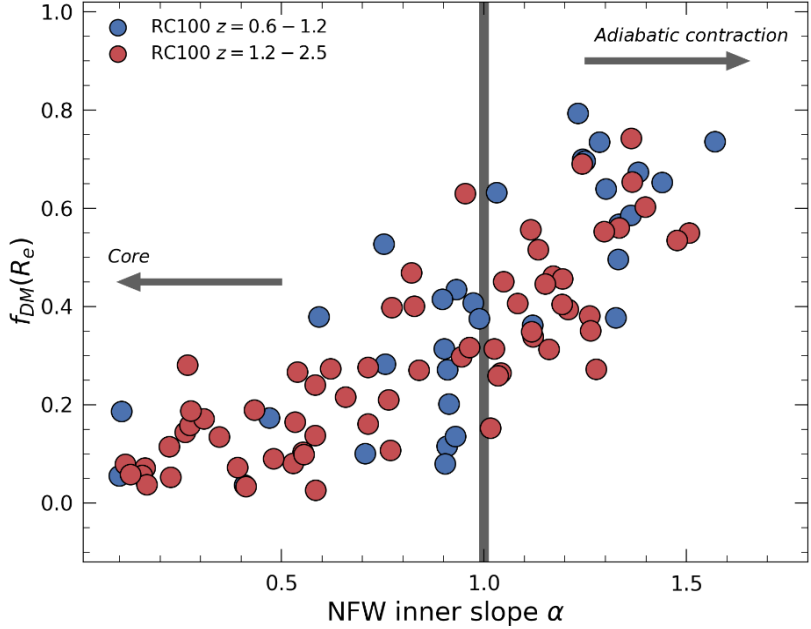

*Figure 8: Effective inner slope ($\rho \propto r^{-\alpha}$) matching the inferred DM within $R_e$ assuming the SMHM relation holds for the total halo mass (Moster et al. 2018). Blue (red) circles show $z = 0.6 - 1.2$ ($z = 1.2 - 2.5$) RC100 galaxies. The vertical grey line shows the NFW slope ($\alpha = 1$), with smaller values representing cored distributions and higher representing effective adiabatic contraction. Low dark matter fractions gravitate towards smaller inner slopes, with negligible fractions $f_{DM} \lesssim 0.1$ matching with flat cores. Cores are preferential for high-z galaxies, with roughly 25% (18) of the high-z galaxies having a slope less than 0.5 and only 10% (3) of the low-z galaxies.*

### 4.6 Adaptive optics (AO) assisted subset of RC100

***Better constraint on bulges.*** Determining bulge masses in our high-z data sets is a difficult task, given that the assumed bulge radius of 1 kpc is comparable to the angular resolution of HST imaging and AO-assisted IFU data with FWHM ~ 1.5 kpc. However, the inner rotation and dispersion curves are more sensitive to the presence of a bulge, with a specific feature (central rise) in the observed velocity dispersion due to sharp gradients in the rotation velocity. Higher resolution gives better constraints on the determining the bulge mass, as such features would become less effected by other components. At our best angular resolution, the AO-assisted data separates this feature most distinctively.

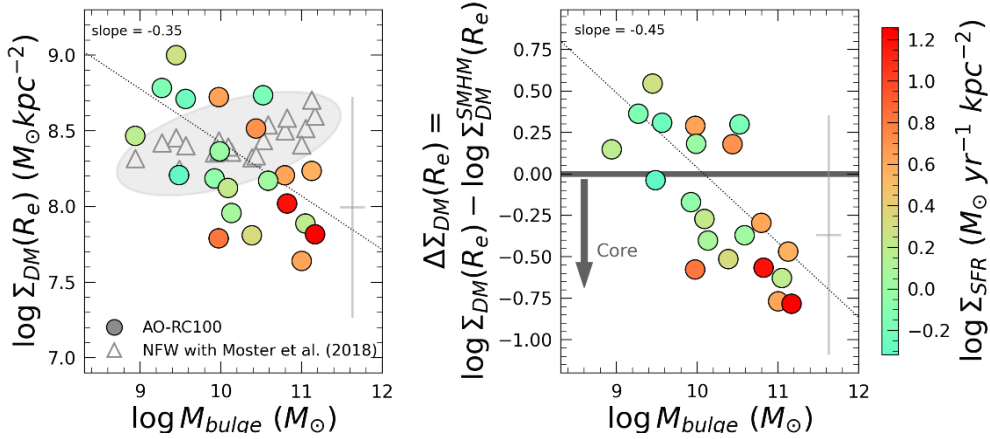

*Figure 9: RC100 galaxies observed with adaptive-optics (PSF FWHM ~0.1-0.3"). Left: DM surface density as a function of bulge mass, with predicted values assuming the SMHM relation of Moster et al. (2018) in grey triangles. Right: Ratio of inferred DM surface densities to values from assuming the SMHM relation ("offset"). In both panels color coding shows star-formation surface density. Both panels show an anti-correlation with a Spearman coefficient of -0.52 (left) and -0.71 (right). Galaxies showing a deficit in DM tend to have higher SFR surface densities.*

*Massive bulges drive DM deficit.* This AO assisted subset of RC100 contains 22 galaxies (Förster Schreiber et al. 2018), for which the dependence of the DM surface density on bulge mass becomes more evident. Figure 9 shows $\Sigma_{DM}(R_e)$ (left) and the offset $\Delta \log \Sigma_{DM}(R_e)$ (right) vs. $M_{bulge}$, color coded by $\Sigma_{SFR}$. A clearer anti-correlation can be seen, with higher bulge mass higher SFR resulting in much lower DM content (Spearman coefficient = -0.52, -0.71 respectively, both $p < 0.05$). As in the previous section, we find that $M_{bulge} \approx 10^{10}$ M$_\odot$ splits the subset reasonably well, with a lower offset (DM "deficit") above it and high offset (DM abundance) above it (see Table 2). We fit the left panel of Figure 9 with a power law of the form $\log\left(\frac{\Sigma_{DM}(R_e)}{M_\odot \text{kpc}^{-2}}\right) = A + B \log\left(\frac{M_{bulge}}{10^{10} M_\odot}\right)$, finding $A = 8.42 \pm 0.07$ and $B = -0.35 \pm 0.11$.

*Removed DM is a fraction of bulge mass.* The amount of DM driven out of the inner regions of the galaxy is dependent on the bulge mass. Among these galaxies with a deficit of DM, Figure 10 plots the amount of removed dark matter when comparing to the predicted SMHM relation, $M_{DM,removed}(R_e) = M_{DM}^{SMHM}(R_e) - M_{DM}(R_e)$, as a function of $M_{bulge}$. On average, these galaxies removed a DM mass equivalent to $\approx 25\%$ of the bulge mass, and none have removed more than the mass of the bulge. Our results are in broad agreement with G20, which have shown that the removed DM mass is $30\% \pm 10\%$ of the bugle mass. It is plausible to connect the formation of the DM core with the creation of a massive central bulge of $M_{bulge} \gtrsim 10^{10}$ M$_\odot$ already at $z \sim 2$, for which the amount of removed DM is correlated with the mass of the bulge. These galaxies also have $B/T$ ratios of $\approx 0.5$ on average. A plausible explanation is due to radial inflows transporting baryons to the center and heat the dark matter via dynamical friction, increasing its kinetic energy and making it less bound gravitationally. If this is followed by strong feedback (most likely AGN-driven) this could lead to driving DM from the disk scale to the out halo (El Zant et al. 2001; Governato et al. 2012; Freundlich et al. 2020; Dekel et al. 2021; Li et al. 2022; Ogiya & Nagai 2022). However, given the large uncertainties (due to comparing two highly uncertain values) this relation is not strongly constrained.

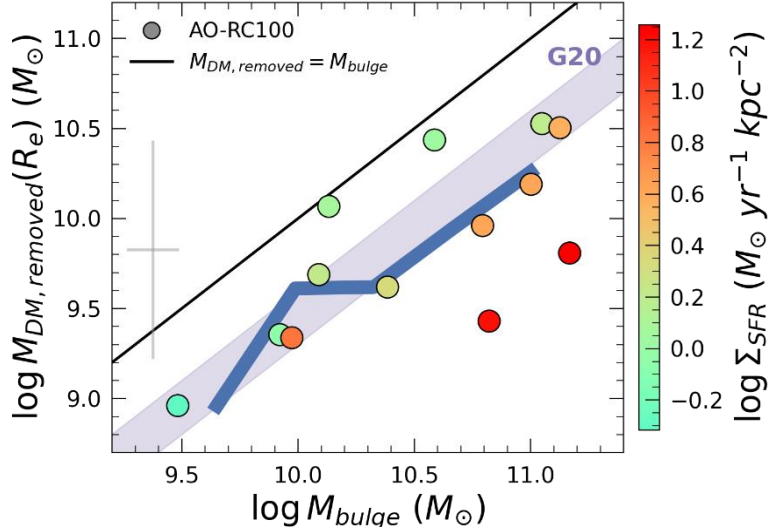

*Figure 10: AO-RC100 galaxies with a DM deficit, showing the amount of removed dark matter within $R_e$ compared to predicted SMHM relations, $M_{DM,removed}(R_e) = M_{DM}^{SMHM}(R_e) - M_{DM}(R_e)$, as a function of bulge mass, with the mean uncertainty shown as the large grey cross. Color bar shows the SFR surface density. No galaxies have removed DM matter mass greater than their bulge mass (solid line). The blue line is showing the moving average, approximately matching a removal of 25% of the bulge mass, showing similar values to the 20%-40% range presented in G20 (purple shaded region).*

## 5. Summary

We have presented kinematic forward-modeling for $H\alpha$ and $CO$ extended rotation curves ($\sim 2\,R_e$) of a hundred main-sequence massive star-forming galaxies (MS SFGs) at redshifts $z = 0.6 - 2.5$. Our three fitting procedures were applied individually for each RC and the best-fit values agree very well with one another (Pearson correlation coefficients of $\geq 0.72$ and up to $0.94$, see Appendix C). These provides more robust estimates for the best-fitting values of the mass modelling for each galaxy, namely $f_{DM}(R_e), M_{baryon}, R_e, \sigma_0$.

Our most prominent result is that at $z \sim 2$ SFGs can become heavily baryon dominated within $R_e$, with half of all disks being less than "maximal" ($f_{DM}(R_e) < 0.28$), while some can become dark matter dominated (up to $f_{DM}(R_e) \approx 0.80$). The median DM fractions for RC100 decrease with redshift as $< f_{DM}(R_e) > = 0.75^{\pm 0.23}\,(1+z)^{-1.07\pm 0.33}$, from $0.38$ at $z = 0.85$ to $0.17$ at $z = 2.44$. Such low DM fractions can be attributed to flat ("cored") DM distributions. In addition, we find a strong correlation between $f_{DM}(R_e)$ and the baryon surface density within $R_e$, $\Sigma_{baryon}(R_e)$, agreeing well with similar studies on high redshift SFGs. We note that these baryon-dominated SFGs at $z \sim 2$ are more similar to early type galaxies (ETG) in the Local Universe, making them their plausible progenitors.

The kinematically inferred DM within $R_e$ can be much lower (but sometimes higher) than precited by stellar-mass halo-mass (SMHM) relations (e.g., Moster et al. 2018). In RC100, the "deficit" of dark matter within $R_e$ is more frequent at $z \sim 2$, and at high baryon surface density ($\Sigma_{baryon} > 10^{8.5}$ M$_\odot$kpc$^{-2}$) and massive bulges ($M_{bulge} > 10^{10}$ M$_\odot$). We also find larger DM deficits at higher SFR surface density ($\Sigma_{SFR} > 10^{0.25}$ M$_\odot$kpc$^{-2}$yr$^{-1}$), with half of all RC100 galaxies having more than 50% of their DM within $R_e$ removed. There is a strong correlation of the DM deficit with $M_{bulge}$ and $\Sigma_{SFR}$, seen clearly for the adaptive-optics assisted subset of RC100. The removed DM mass for these AO-RC100 galaxies is on average 25% of the bulge mass, with none removing more DM mass than their bulge's mass.

In massive galaxies AGN feedback is expected to dominate over stellar feedback, as the central BH mass scales with the mass of the system, which can lead to a change in the shape of the DM distribution on large scales. Galaxies with bulge masses above $10^{10}$ M$_\odot$ are much more likely to have a broad nuclear component indicative of an AGN feedback (~30% − 50% of galaxies, Genzel et al. 2014), and higher SFR surface density a higher incidence of SF-driven broad components (~15% − 25% of galaxies, Förster Schreiber et al. 2019). Strong AGN driven outflows or bursty star formation can remove large amounts of baryons from the inner regions of the galaxy, leading to a formation of a core in the DM halo as the particle adjust to the change in the gravitational potential (De Souza et al. 2011; Ogiya and Mori 2011; Governato et al. 2012; Li et al. 2022). Dynamical friction due to radial inflows may pre-heat the DM particles, making them more reactive to the feedback (Freundlich et al. 2020; Dekel et al. 2021; Ogiya & Nagai 2022). Most of these models currently suggest that cores only form if the galaxy suddenly loses the majority of its gas content, which, if significant, can quench star formation and lead the galaxy to become an ETG in its later stages of evolution.

Our increased statistics and deep high-quality observations allow us to strengthen our previous conclusions (Genzel et al. 2017, 2020; Price et al. 2021) of baryon dominated disks at $z\sim2$, linking the DM deficit to cored distributions and high $\Sigma_{baryon}$, $M_{bulge}$ and $\Sigma_{SFR}$, with AGN and SF-driven feedback as possible mechanism driving DM outwards. Future observations at higher resolution are important in order to better constrain the inner structures of these galaxies, including the bulges, as well as their clumpy structure, and in assessing the importance of non-circular motions that could signal efficient radial inflow of matter rapidly accumulating at the galaxy centers.

## Acknowledgments

We thank our colleagues at MPE, ESO-Garching and ESO-Paranal, LBT and IRAM, and members of the 3D-HST, SINFONI/SINS & zC-SINF and KMOS/KMOS$^{3D}$ teams, who have contributed to, helped, or otherwise supported these observations and their analysis. This paper is based on observations collected at the European Organisation for Astronomical Research in the Southern Hemisphere under ESO Programmes 074.A-9011, 075.A-0466, 076.A-0527, 078.A-0600, 079.A-0341, 080.A-0330, 080.A-0339, 080.A-0635, 081.A-0672, 082.A-0396, 088.A-0202, 088.A-0209, 090.A-0516, 091.A-0126, 092.A-0082, 092.A-0091, 093.A-0079, 093.A-0110, 093.A-0233, 094.A-0217, 094.A-0568, 095.A-0047, 096.A-0025, 097.A-0028, 097.B-0065, 097.A-0353, 098.A-0045, 099.A-0013, 099.B-0275, 100.A-0039, 100.A-0361, 101.A-0022, 102.B-0062, 102.B-0087, and 183.A-0781. Also based on observations carried out with the IRAM NOEMA Interferometer. IRAM is supported by INSU/CNRS (France), MPG (Germany) and IGN (Spain). This paper makes use of the following ALMA data: ADS/JAO.ALMA#2013.1.00952.S,

ADS/JAO.ALMA#2016.1.00406.S, ADS/JAO.ALMA#2019.1.00640.S, and ADS/JAO.ALMA#2019.1.01362.S. ALMA is a partnership of ESO (representing its member states), NSF (USA) and NINS (Japan), together with NRC (Canada), MOST and ASIAA (Taiwan), and KASI (Republic of Korea), in cooperation with the Republic of Chile. The Joint ALMA Observatory is operated by ESO, AUI/NRAO and NAOJ. This paper also uses observations obtained at the Large Binocular Telescope (LBT). The LBT is an international collaboration among institutions in the United States, Italy and Germany. LBT Corporation partners are: LBT Beteiligungsgesellschaft, Germany, representing the Max-Planck Society, the Astrophysical Institute Potsdam, and Heidelberg University; The University of Arizona on behalf of the Arizona university system; Istituto Nazionale di Astrofisica, Italy; The Ohio State University, and The Research Corporation, on behalf of The University of Notre Dame, University of Minnesota and University of Virginia. We thank the German Science Foundation (DFG) for support via German-Israel Project (DIP) grant STE/1869-2 GE 625/17-1. AS is supported by research grants from the Center for Computational Astrophysics (CCA) of the Flatiron Institute, and the Mathematics and Science Division of the Simons Foundation, USA. AD has been partly supported by grants ISF 861/20 and DIP 030-9111. EW acknowledges support by the Australian Research Council Centre of Excellence for All Sky Astrophysics in 3 Dimensions (ASTRO 3D), through project number CE170100013. HÜ gratefully acknowledges support by the Isaac Newton Trust and by the Kavli Foundation through a Newton-Kavli Junior Fellowship.

*Software.* Astropy (Astropy Collaboration et al. 2013, 2018), emcee (Foreman-Mackey et al. 2013), corner (Foreman-Mackey 2016), Matplotlib (Hunter 2007), MPFIT (Markwardt 2009), Numpy (Van Der Walt et al. 2011; Harris et al. 2020), Pandas (McKinney 2010; The pandas development team 2020), Scipy (Virtanen et al. 2020).

## Appendix A

The mass model consists of a thick turbulent disk with a constant velocity dispersion, a bulge and a dark matter halo. The disk and bulge are modeled as de-projected Sérsic profiles with total baryonic mass ($M_{baryon}$) and bulge-to-total ratio $B/T = M_{bulge}/M_{baryon}$. Light traces mass with a constant $M/L$ ratio and the bulge does not emit light in the kinematic tracers. The projected surface density for the disk and bulge follows a Sérsic profile (Binney & Tremaine 1987):

$$[5] \quad \Sigma_{Sérsic}(r) = \Sigma_0 e^{-b\left(\frac{r}{R_e}\right)^{1/n_s}}$$

where $R_e$ is the effective radius, $n_s$ is the Sérsic index ($n_s = 1$ for an exponential disk), $\Sigma_0 = \frac{M}{2\pi R_e^2} \frac{b^{2n_s}}{n_s \Gamma(2n_s)}$ is the characteristic density, and $b$ is given by the implicit equation $\frac{1}{2}\Gamma(2n_s) = \gamma(2n_s, b)$ ($\Gamma$ and $\gamma$ are the gamma function and the lower incomplete gamma function, respectively). We assume the bulge to be spherical with a Sérsic index of $n_{s,bulge} = 4$ and projected effective radius of $R_{e,bulge} = 1$ kpc. The disk is taken as a flattened spheroid with vertical to radial extent ratio $q_{0,disk}$ ($0.2 - 0.25$ for disks at high-z), Sérsic index $n_{s,disk}$ (hereafter $n_s$), and effective radius $R_{e,disk}$ (hereafter $R_e$). The circular velocity from the de-projected density distribution is given by Noordermeer (2008).

The dark matter halo is assumed to be spherically symmetric and follow a NFW density profile (Navarro, Frenk & White 1996) with a total mass $M_{vir}$ and concentration parameter $c = r_{vir}/r_s$. The virial radius is given as the radius in which the mean density is $200\rho_c$. The density profile is given by:

$$[6] \quad \rho_{DM}(r) = \frac{\rho_0}{(r/r_s)(1+r/r_s)^2}$$

were $\rho_0$ is a characteristic density and $r_s$ is the scale radius. The circular velocity is given by $V_{DM}(r) = \sqrt{\frac{GM_{DM}(r)}{r}}$. Adiabatic contraction of the halo due to the formation of the galaxy is considered only when stated so, otherwise a regular NFW profile is used. As will be defined later, the DM fraction can be used as a free parameter for the mass of the halo instead of $M_{vir}$.

The circular velocity due to the gravitational potentials of all three mass components is:

$$[7] \quad V_{circ}^2(r) = V_{disk}^2(r) + V_{bulge}^2(r) + V_{DM}^2(r)$$

where $V_{disk}(r)$, $V_{bulge}(r)$, $V_{halo}(r)$ are the circular velocities of the disk, bulge and DM halo. Additionally, pressure gradients due to turbulent motions lead to a reduction in the circular velocity. In general, the correction for an isotropic self-gravitating disk with velocity dispersion $\sigma$ is in the form of $V_{rot}^2(r) = V_{circ}^2(r) + 2\sigma^2 \frac{d\ln\Sigma}{d\ln r}$ (Burkert et al. 2010). We take the correction for an exponential disk with a constant velocity dispersion $\sigma_0$:

$$[8] \quad V_{rot}^2(r) = V_{circ}^2(r) - 3.36\sigma_0^2 \left(\frac{r}{R_e}\right)$$

with $V_{disk}(r), V_{bulge}(r), V_{halo}(r)$ the velocity at radius $r$ due to the disk, bulge and halo respectively. The correction is always negative and proportional to $(r/R_e)$, thus increasing in the outer rotation curve. If the velocity dispersion is high enough compared to the peak rotation curve (i.e., low $V/\sigma_0$), the RC will start decreasing significantly close to $\sim R_e$.

## Appendix A.1

In Methods A and B described in section 3 we construct fully-forward 4D hypercube models used, which are outlined here and the reader is referred to G20 and Price et al. (2021) for details. We create a 4D hypercube for each galaxy based on the parameter values, by generating a velocity Gaussian at each point on the 3D cube oriented with the plane of the sky $(x_{sky}, y_{sky}, z_{sky})$, given the galaxy's inclination and position angle (PA). The velocity Gaussian is centered on the intrinsic projected rotation velocity ($V_{rot}$) with intrinsic dispersion ($\sigma_0$) and flux assuming that the disk mid-plane line intensity follows the disk mass distribution (i.e., constant $M/L$). We assume a Gaussian vertical flux distribution reflecting the assumed disk intrinsic axis ratio. The 4D hypercube is collapsed along the line-of-sight ($z_{sky}$), resulting in a model-intrinsic 3D cube $(x_{sky}, y_{sky}, V_{los})$. We convolve this cube with both the spatial PSF and the spectral line spread function (LSF) to yield a model 3D cube that includes all observational effects. We then perform an aperture extraction on this 3D model cube – using the same slit/apertures and Gaussian fitting to measure $V_{obs}$ and $\sigma_{obs}$ as used on the observational data – to obtain 1D model rotation and dispersion curves along the galaxy major axis. Finally, since the observed 1D dispersions have been corrected for the instrumental resolution, the extracted model dispersions are also corrected by subtracting the instrument dispersion in quadrature.

## Appendix A.2

This section further explains the procedure we refer to as "beam-projection" beam smearing. This is a *Python* implementation of a 2D convolution on the galactic plane of the rotation velocities. The intrinsic velocities are calculated from the same mass modelling (disk, bulge and DM halo) described in Section 2 and Appendix A.

We construct a two-dimensional velocity grid on the galactic plane ($x_{gal}, y_{gal}$), so that the cylindrical radius is $r = \sqrt{x_{gal}^2 + y_{gal}^2}$, with the spatial resolution chosen to be much smaller than the beam size (~ 1/50) and uniformly sampled. This ensures that the grid is sampled finely enough so that numerical errors are negligible. We position the coordinates so that the major-axis lies on the x-axis and the minor-axis on the y-axis of the galaxy, so that projected rotation velocities along the line-of-sight are given by their coordinates on the galaxy's plane, with $V_{LOS}(r) = V_{rot}(r) \cos\left(\frac{x_{gal}}{r}\right)$. We convolve the velocity-grid with a Gaussian PSF weighted by the flux-intensity assuming a mass-to-light ratio of 1. Only the disk contributed directly to the light weighting, as the bulge is assumed to make no contribution to the line emission. Inclination is crucial, as it not only lowers the observed velocity by a factor of $\sin(i)$, it changes the way the beam interacts with the galactic plane. With $i = 0°$ being face-on, the projected area of the beam on the galaxy's plane increases with increasing inclination. As the minor-axis coincides with the galaxy y-axis, this means that the y coordinate of the PSF is elongated as the inclination increases. In the galactic coordinates, the beam projects an ellipse with its FWHM elongated along the y-axis by a factor of $y_{FWHM,gal} = y_{FWHM}/\cos(i)$. The observer PSF is taken to be Gaussian with a $\sigma = FWHM/\left(2\sqrt{2\ln 2}\right)$, so on the galactic plane the PSF would take the form of:

$$PSF_{gal} = \exp\left\{-\frac{x_{gal}^2}{2\sigma_{beam}^2} - \frac{y_{gal}^2}{2(\sigma_{beam}/\cos(i))^2}\right\}$$

where $PSF_{gal}$ is the projected Gaussian beam on the galactic coordinates. Larger inclination would change the contribution of similar $y_{gal}$ values to the convolution, allowing farther regions to become relevant. As we approach an edge-on geometry, $i \rightarrow 90°$, the denominator on the last term diverges, $\sigma_{beam}/\cos(i) \rightarrow \infty$, and the $PSF_{gal}$ becomes effectively dependent on one coordinate alone. For a face-on galaxy, $i = 0°$, we recover the circular PSF (although in this case the observed tangential line-of-sight will be formally zero).

## Appendix B

Following our three fitting methods, we average the best-fit values with their combined uncertainties. The values are listed in the following table, in column order: (1) Galaxy name, (2) redshift, (3) PSF FWHM, (4) integration time, (5) logarithmic offset in SFR from the MS at the stellar mass and redshift of the galaxy, (6) stellar mass, (7) total baryonic mass, (8) bulge mass, (9) disk effective radius, (10) DM fraction within $R_e$, (11) circular velocity at $R_e$, (12) intrinsic velocity dispersion, (13) halo mass assuming the SMHM, (14) SFR surface density within $R_e$, (15) kinematically inferred DM surface density.

Table 3: RC100 galaxy best-fit parameters and characteristics

| | Galaxy | $z$ | FWHM ["] | $T_{int}$ [hrs] | $\delta \log (SFR/SFR(MS))$ | $\log (M_\star /M_\odot)$ | $\log (M_{baryon} /M_\odot)$ | $\log (M_{bulge} /M_\odot)$ | $R_e/kpc$ | $f_{DM}(<R_e)$ | $V_c(R_e)$ [km/s] | $\sigma_0$ [km/s] | $\log (M_{vir}^{SMHM} /M_\odot)$ | $\log (\Sigma_{SFR}/M_\odot yr^{-1}kpc^{-2})$ | $\log \Sigma_{DM}(<R_e) /M_\odot kpc^{-2}$ |
|---|---|---|---|---|---|---|---|---|---|---|---|---|---|---|---|
| (1) | U4 38210 | 0.61 | 0.75 | 4.9 | 0.14 | 10.85 | $10.99^{\pm 0.13}$ | $10.44^{\pm 0.19}$ | $8.10^{\pm 1.53}$ | $0.44^{\pm 0.13}$ | $240^{\pm 48}$ | $44^{\pm 7}$ | 12.44 | -1.14 | $8.36^{\pm 0.54}$ |
| (2) | U4 20688 | 0.63 | 0.7 | 4.9 | 0.16 | 11.0 | $10.81^{\pm 0.11}$ | $10.62^{\pm 0.12}$ | $9.57^{\pm 1.65}$ | $0.67^{\pm 0.13}$ | $275^{\pm 55}$ | $16^{\pm 6}$ | 12.2 | -1.18 | $8.59^{\pm 0.44}$ |
| (3) | EGS3 10098 | 0.66 | 0.77 | 21.0 | 0.49 | 11.11 | $10.93^{\pm 0.11}$ | $10.71^{\pm 0.13}$ | $2.95^{\pm 0.92}$ | $0.14^{\pm 0.12}$ | $271^{\pm 54}$ | $29^{\pm 12}$ | 12.36 | 0.28 | $8.40^{\pm 1.20}$ |
| (4) | U3 21388 | 0.67 | 0.54 | 8.2 | -0.37 | 10.76 | $10.64^{\pm 0.18}$ | $9.33^{\pm 0.33}$ | $7.81^{\pm 1.28}$ | $0.70^{\pm 0.13}$ | $205^{\pm 41}$ | $53^{\pm 7}$ | 12.02 | -1.61 | $8.44^{\pm 0.41}$ |
| (5) | EGS4 21351 | 0.73 | 0.78 | 19.0 | 0.72 | 10.94 | $10.61^{\pm 0.14}$ | $10.27^{\pm 0.17}$ | $4.16^{\pm 1.66}$ | $0.06^{\pm 0.08}$ | $159^{\pm 32}$ | $17^{\pm 6}$ | 11.89 | 0.17 | $7.40^{\pm 0.87}$ |
| (6) | GS4 00984 | 0.73 | 0.52 | 8.1 | 0.04 | 9.48 | $9.72^{\pm 0.21}$ | $9.02^{\pm 0.29}$ | $5.25^{\pm 1.67}$ | $0.79^{\pm 0.14}$ | $116^{\pm 23}$ | $25^{\pm 5}$ | 11.35 | -1.47 | $8.18^{\pm 0.55}$ |
| (7) | EGS4 11261 | 0.75 | 0.86 | 24.0 | 0.56 | 11.25 | $11.12^{\pm 0.08}$ | $10.81^{\pm 0.12}$ | $5.02^{\pm 1.22}$ | $0.20^{\pm 0.10}$ | $294^{\pm 59}$ | $22^{\pm 7}$ | 12.58 | 0.03 | $8.41^{\pm 0.80}$ |
| (8) | GS4 13143 | 0.76 | 0.52 | 8.9 | 0.33 | 9.82 | $10.33^{\pm 0.11}$ | $10.18^{\pm 0.13}$ | $5.51^{\pm 1.26}$ | $0.41^{\pm 0.13}$ | $147^{\pm 29}$ | $22^{\pm 7}$ | 11.61 | -0.9 | $8.07^{\pm 0.54}$ |
| (9) | U3 08493 | 0.79 | 0.56 | 9.0 | 0.3 | 10.77 | $11.08^{\pm 0.11}$ | $10.48^{\pm 0.19}$ | $3.79^{\pm 1.01}$ | $0.10^{\pm 0.11}$ | $267^{\pm 53}$ | $16^{\pm 10}$ | 12.5 | -0.18 | $8.14^{\pm 1.00}$ |
| (10) | COS4 05656 | 0.8 | 0.5 | 6.2 | 0.93 | 10.68 | $10.61^{\pm 0.10}$ | $10.32^{\pm 0.13}$ | $4.99^{\pm 1.39}$ | $0.27^{\pm 0.12}$ | $181^{\pm 36}$ | $24^{\pm 5}$ | 11.8 | 0.2 | $8.12^{\pm 0.74}$ |
| (11) | COS4 05925 | 0.8 | 0.54 | 6.2 | 0.32 | 10.63 | $10.95^{\pm 0.15}$ | $10.55^{\pm 0.18}$ | $5.72^{\pm 1.08}$ | $0.36^{\pm 0.17}$ | $257^{\pm 51}$ | $35^{\pm 11}$ | 12.25 | -0.53 | $8.49^{\pm 0.78}$ |
| (12) | U3 03856 | 0.8 | 0.54 | 4.4 | -0.08 | 10.59 | $10.62^{\pm 0.17}$ | $10.02^{\pm 0.23}$ | $7.54^{\pm 1.55}$ | $0.64^{\pm 0.20}$ | $212^{\pm 42}$ | $25^{\pm 8}$ | 11.93 | -1.18 | $8.45^{\pm 0.59}$ |
| (13) | U3 05138 | 0.81 | 0.73 | 16.2 | -0.3 | 10.2 | $10.46^{\pm 0.11}$ | $10.16^{\pm 0.13}$ | $7.60^{\pm 1.46}$ | $0.38^{\pm 0.16}$ | $136^{\pm 27}$ | $29^{\pm 6}$ | 11.77 | -1.48 | $7.83^{\pm 0.59}$ |
| (14) | U4 42393 | 0.82 | 0.7 | 4.9 | 0.82 | 10.93 | $11.15^{\pm 0.10}$ | $10.63^{\pm 0.17}$ | $2.61^{\pm 0.73}$ | $0.08^{\pm 0.07}$ | $335^{\pm 67}$ | $64^{\pm 11}$ | 12.52 | 0.78 | $8.41^{\pm 1.17}$ |
| (15) | GS4 03228 | 0.82 | 0.53 | 8.1 | 0.48 | 9.49 | $10.08^{\pm 0.17}$ | $9.99^{\pm 0.18}$ | $6.79^{\pm 1.46}$ | $0.70^{\pm 0.13}$ | $142^{\pm 28}$ | $13^{\pm 7}$ | 11.44 | -1.16 | $8.18^{\pm 0.44}$ |
| (16) | GS4 32976 | 0.83 | 0.52 | 6.9 | 0.22 | 10.37 | $10.70^{\pm 0.09}$ | $10.65^{\pm 0.10}$ | $7.73^{\pm 1.45}$ | $0.57^{\pm 0.09}$ | $235^{\pm 47}$ | $43^{\pm 6}$ | 11.93 | -0.93 | $8.48^{\pm 0.46}$ |
| (17) | GS3 23200 | 0.83 | 0.46 | 4.3 | 0.36 | 10.07 | $10.25^{\pm 0.18}$ | $8.73^{\pm 0.94}$ | $6.72^{\pm 1.49}$ | $0.63^{\pm 0.16}$ | $127^{\pm 25}$ | $33^{\pm 3}$ | 11.55 | -0.83 | $8.05^{\pm 0.55}$ |
| (18) | COS4 01351 | 0.85 | 0.67 | 12.0 | 0.5 | 10.73 | $11.13^{\pm 0.11}$ | $10.51^{\pm 0.17}$ | $7.52^{\pm 1.41}$ | $0.38^{\pm 0.09}$ | $267^{\pm 53}$ | $58^{\pm 4}$ | 12.42 | -0.49 | $8.42^{\pm 0.50}$ |
| (19) | COS4 06487 | 0.91 | 0.52 | 6.2 | 0.2 | 10.85 | $10.91^{\pm 0.09}$ | $10.75^{\pm 0.11}$ | $4.97^{\pm 1.50}$ | $0.38^{\pm 0.10}$ | $286^{\pm 57}$ | $30^{\pm 7}$ | 12.25 | -0.4 | $8.66^{\pm 0.57}$ |
| (20) | COS3 18434 | 0.91 | 0.54 | 2.0 | 0.45 | 10.77 | $10.73^{\pm 0.10}$ | $10.53^{\pm 0.12}$ | $6.75^{\pm 1.57}$ | $0.50^{\pm 0.14}$ | $232^{\pm 46}$ | $38^{\pm 5}$ | 11.97 | -0.44 | $8.47^{\pm 0.57}$ |
| (21) | U4 19954 | 0.91 | 0.59 | 4.6 | 0.42 | 11.13 | $11.21^{\pm 0.07}$ | $10.93^{\pm 0.10}$ | $8.63^{\pm 1.28}$ | $0.17^{\pm 0.12}$ | $260^{\pm 52}$ | $22^{\pm 11}$ | 12.63 | -0.51 | $8.00^{\pm 0.81}$ |
| (22) | COS3 22796 | 0.91 | 0.65 | 10.8 | -0.16 | 10.32 | $10.61^{\pm 0.19}$ | $9.73^{\pm 0.31}$ | $9.93^{\pm 1.71}$ | $0.53^{\pm 0.19}$ | $152^{\pm 30}$ | $25^{\pm 7}$ | 11.86 | -1.5 | $7.96^{\pm 0.68}$ |
| (23) | U4 13849 | 0.91 | 0.52 | 11.8 | 0.15 | 10.71 | $10.88^{\pm 0.12}$ | $10.42^{\pm 0.16}$ | $6.79^{\pm 1.35}$ | $0.31^{\pm 0.13}$ | $213^{\pm 43}$ | $29^{\pm 5}$ | 12.18 | -0.76 | $8.19^{\pm 0.59}$ |
| (24) | U4 22199 | 0.91 | 0.59 | 4.6 | 0.26 | 10.6 | $10.46^{\pm 0.06}$ | $10.44^{\pm 0.07}$ | $4.93^{\pm 1.19}$ | $0.02^{\pm 0.04}$ | $117^{\pm 23}$ | $35^{\pm 5}$ | 11.77 | -0.38 | $6.71^{\pm 0.62}$ |
| (25) | U3 15226 | 0.92 | 0.75 | 19.8 | 0.05 | 11.11 | $10.72^{\pm 0.10}$ | $10.46^{\pm 0.13}$ | $5.79^{\pm 1.28}$ | $0.28^{\pm 0.12}$ | $182^{\pm 36}$ | $30^{\pm 6}$ | 12.07 | -0.53 | $8.08^{\pm 0.61}$ |
| (26) | GS4 05881 | 0.99 | 0.57 | 13.9 | 0.46 | 9.78 | $10.21^{\pm 0.13}$ | $10.14^{\pm 0.14}$ | $5.71^{\pm 1.41}$ | $0.74^{\pm 0.11}$ | $193^{\pm 39}$ | $61^{\pm 5}$ | 11.52 | -0.73 | $8.55^{\pm 0.49}$ |
| (27) | GS3 30840 | 1.02 | 0.4 | 4.2 | 0.33 | 10.31 | $10.51^{\pm 0.13}$ | $10.21^{\pm 0.15}$ | $5.82^{\pm 1.64}$ | $0.65^{\pm 0.11}$ | $207^{\pm 41}$ | $38^{\pm 5}$ | 11.72 | -0.5 | $8.55^{\pm 0.53}$ |
| (28) | GS4 29973 | 1.02 | 0.49 | 5.7 | 0.05 | 9.59 | $9.90^{\pm 0.20}$ | $9.50^{\pm 0.23}$ | $5.09^{\pm 1.54}$ | $0.74^{\pm 0.12}$ | $135^{\pm 27}$ | $24^{\pm 6}$ | 11.41 | -1.19 | $8.29^{\pm 0.54}$ |
| (29) | GS3 27242 | 1.02 | 0.4 | 8.3 | 0.07 | 10.64 | $10.85^{\pm 0.10}$ | $10.45^{\pm 0.14}$ | $2.85^{\pm 0.88}$ | $0.12^{\pm 0.08}$ | $259^{\pm 52}$ | $12^{\pm 5}$ | 12.15 | -0.02 | $8.31^{\pm 0.86}$ |
| (30) | COS3 16954 | 1.03 | 0.9 | 4.0 | 0.55 | 10.74 | $10.90^{\pm 0.10}$ | $10.60^{\pm 0.13}$ | $8.26^{\pm 1.47}$ | $0.59^{\pm 0.10}$ | $269^{\pm 54}$ | $51^{\pm 5}$ | 12.06 | -0.33 | $8.58^{\pm 0.45}$ |
| (31) | COS3 04796 | 1.03 | 0.7 | 9.8 | 0.25 | 10.8 | $11.18^{\pm 0.11}$ | $10.43^{\pm 0.20}$ | $9.60^{\pm 1.58}$ | $0.42^{\pm 0.11}$ | $266^{\pm 53}$ | $14^{\pm 6}$ | 12.46 | -0.75 | $8.36^{\pm 0.49}$ |
| (32) | EGS 13035123 | 1.12 | 0.66 | 56.0 | 0.31 | 11.18 | $11.08^{\pm 0.09}$ | $10.38^{\pm 0.15}$ | $10.19^{\pm 1.63}$ | $0.19^{\pm 0.13}$ | $196^{\pm 39}$ | $16^{\pm 3}$ | 12.33 | -0.41 | $7.72^{\pm 0.73}$ |
| (33) | EGS 13004291 | 1.2 | 0.51 | 38.0 | 1.25 | 10.97 | $11.11^{\pm 0.06}$ | $10.90^{\pm 0.09}$ | $3.65^{\pm 1.02}$ | $0.04^{\pm 0.06}$ | $284^{\pm 57}$ | $54^{\pm 6}$ | 11.95 | 1.18 | $7.78^{\pm 0.93}$ |
| (34) | EGS 13003805 | 1.23 | 0.62 | 20.0 | 0.56 | 11.23 | $11.46^{\pm 0.07}$ | $10.93^{\pm 0.15}$ | $5.94^{\pm 1.18}$ | $0.05^{\pm 0.08}$ | $321^{\pm 64}$ | $29^{\pm 9}$ | 12.71 | 0.26 | $7.83^{\pm 0.96}$ |
| (35) | GS4 30432 | 1.31 | 0.55 | 12.0 | -0.21 | 10.7 | $10.66^{\pm 0.11}$ | $10.50^{\pm 0.12}$ | $7.21^{\pm 1.62}$ | $0.52^{\pm 0.12}$ | $208^{\pm 42}$ | $38^{\pm 5}$ | 11.95 | -0.91 | $8.36^{\pm 0.53}$ |
| (36) | GS4 14356 | 1.36 | 0.65 | 2.0 | 0.41 | 10.57 | $10.72^{\pm 0.19}$ | $8.72^{\pm 1.59}$ | $5.59^{\pm 1.26}$ | $0.46^{\pm 0.23}$ | $193^{\pm 39}$ | $26^{\pm 15}$ | 11.86 | -0.08 | $8.35^{\pm 0.74}$ |
| (37) | EGS4 38153 | 1.36 | 0.6 | 30.0 | 0.25 | 10.44 | $11.16^{\pm 0.16}$ | $10.37^{\pm 0.27}$ | $5.11^{\pm 1.28}$ | $0.35^{\pm 0.14}$ | $314^{\pm 63}$ | $57^{\pm 12}$ | 12.24 | -0.02 | $8.70^{\pm 0.70}$ |
| (38) | GS4 27441 | 1.38 | 0.54 | 13.8 | 0.35 | 10.82 | $11.06^{\pm 0.17}$ | $10.44^{\pm 0.25}$ | $7.21^{\pm 1.55}$ | $0.46^{\pm 0.15}$ | $271^{\pm 54}$ | $68^{\pm 9}$ | 12.22 | -0.26 | $8.54^{\pm 0.66}$ |
| (39) | EGS4 24985 | 1.4 | 0.85 | 80.0 | 0.37 | 10.9 | $11.03^{\pm 0.10}$ | $10.59^{\pm 0.15}$ | $4.64^{\pm 1.12}$ | $0.31^{\pm 0.12}$ | $288^{\pm 58}$ | $41^{\pm 11}$ | 12.25 | 0.17 | $8.62^{\pm 0.58}$ |
| (40) | U4 27256 | 1.42 | 0.62 | 5.4 | 0.12 | 10.49 | $10.83^{\pm 0.13}$ | $10.11^{\pm 0.23}$ | $6.37^{\pm 1.42}$ | $0.14^{\pm 0.10}$ | $168^{\pm 34}$ | $20^{\pm 8}$ | 11.98 | -0.48 | $7.68^{\pm 0.82}$ |

| | Galaxy | $z$ | FWHM ["] | $T_{int}$ [hrs] | $\delta \log(SFR/SFR(MS))$ | $\log(M_\star/M_\odot)$ | $\log(M_{baryon}/M_\odot)$ | $\log(M_{bulge}/M_\odot)$ | $R_e/kpc$ | $f_{DM}(<R_e)$ | $V_c(R_e)$ [km/s] | $\sigma_0$ [km/s] | $\log(M_{vir}^{SMHM}/M_\odot)$ | $\log(\Sigma_{SFR}/M_\odot yr^{-1} kpc^{-2})$ | $\log \Sigma_{DM}(<R_e)/M_\odot kpc^{-2}$ |
|---|---|---|---|---|---|---|---|---|---|---|---|---|---|---|---|
| (41) | U4 32570 | 1.42 | 0.61 | 20.0 | 0.58 | 10.89 | $10.81^{\pm 0.14}$ | $10.62^{\pm 0.15}$ | $8.01^{\pm 1.66}$ | $0.55^{\pm 0.13}$ | $252^{\pm 50}$ | $89^{\pm 9}$ | 11.96 | -0.07 | $8.51^{\pm 0.60}$ |
| (42) | U4 28358 | 1.42 | 0.49 | 13.0 | 0.26 | 10.96 | $10.98^{\pm 0.11}$ | $10.61^{\pm 0.15}$ | $3.14^{\pm 1.23}$ | $0.15^{\pm 0.11}$ | $294^{\pm 59}$ | $74^{\pm 7}$ | 12.24 | 0.41 | $8.49^{\pm 1.03}$ |
| (43) | zC 403741 | 1.45 | 0.35 | 6.0 | 0.3 | 10.65 | $10.55^{\pm 0.06}$ | $10.39^{\pm 0.09}$ | $3.01^{\pm 0.98}$ | $0.08^{\pm 0.10}$ | $180^{\pm 36}$ | $66^{\pm 5}$ | 11.82 | 0.32 | $7.81^{\pm 1.05}$ |
| (44) | U4 33383 | 1.48 | 0.59 | 7.0 | 0.33 | 9.83 | $10.05^{\pm 0.12}$ | $9.66^{\pm 0.16}$ | $4.98^{\pm 1.34}$ | $0.28^{\pm 0.17}$ | $92^{\pm 18}$ | $48^{\pm 4}$ | 11.44 | -0.53 | $7.55^{\pm 0.75}$ |
| (45) | COS4 10048 | 1.5 | 0.45 | 7.45 | 0.06 | 10.34 | $10.56^{\pm 0.12}$ | $10.04^{\pm 0.18}$ | $7.48^{\pm 1.41}$ | $0.47^{\pm 0.11}$ | $160^{\pm 32}$ | $46^{\pm 5}$ | 11.77 | -0.73 | $8.08^{\pm 0.51}$ |
| (46) | D3a 6397 | 1.5 | 0.29 | 11.0 | 0.53 | 11.08 | $11.24^{\pm 0.06}$ | $11.05^{\pm 0.09}$ | $6.54^{\pm 1.27}$ | $0.13^{\pm 0.10}$ | $225^{\pm 45}$ | $60^{\pm 5}$ | 12.43 | 0.2 | $7.89^{\pm 0.74}$ |
| (47) | U4 31577 | 1.52 | 0.59 | 8.3 | 0.72 | 11.07 | $11.06^{\pm 0.10}$ | $10.56^{\pm 0.16}$ | $6.49^{\pm 1.48}$ | $0.53^{\pm 0.09}$ | $347^{\pm 69}$ | $129^{\pm 8}$ | 12.18 | 0.4 | $8.87^{\pm 0.51}$ |
| (48) | U4 35796 | 1.53 | 0.58 | 8.3 | 0.45 | 10.81 | $11.18^{\pm 0.14}$ | $10.48^{\pm 0.23}$ | $7.29^{\pm 1.15}$ | $0.12^{\pm 0.14}$ | $225^{\pm 45}$ | $29^{\pm 10}$ | 12.29 | -0.1 | $7.77^{\pm 1.02}$ |
| (49) | COS4 13890 | 1.53 | 0.52 | 7.8 | 0.11 | 11.33 | $11.44^{\pm 0.07}$ | $11.18^{\pm 0.11}$ | $7.19^{\pm 1.75}$ | $0.34^{\pm 0.08}$ | $412^{\pm 82}$ | $41^{\pm 5}$ | 12.76 | -0.08 | $8.77^{\pm 0.51}$ |
| (50) | EGS 13011166 | 1.53 | 0.75 | 15.0 | 0.74 | 11.08 | $11.25^{\pm 0.08}$ | $10.99^{\pm 0.11}$ | $6.89^{\pm 1.33}$ | $0.26^{\pm 0.11}$ | $320^{\pm 64}$ | $53^{\pm 7}$ | 12.22 | 0.4 | $8.47^{\pm 0.56}$ |
| (51) | COS4 24012 | 1.53 | 0.59 | 10.0 | 0.42 | 10.25 | $10.48^{\pm 0.12}$ | $10.26^{\pm 0.14}$ | $4.81^{\pm 1.11}$ | $0.19^{\pm 0.12}$ | $139^{\pm 28}$ | $42^{\pm 6}$ | 11.67 | -0.04 | $7.75^{\pm 0.88}$ |
| (52) | COS4 17808 | 1.59 | 0.58 | 7.1 | 0.28 | 10.72 | $10.75^{\pm 0.10}$ | $10.45^{\pm 0.14}$ | $4.46^{\pm 1.55}$ | $0.10^{\pm 0.11}$ | $226^{\pm 45}$ | $26^{\pm 9}$ | 11.94 | 0.17 | $7.94^{\pm 0.92}$ |
| (53) | GK 2540 | 1.61 | 0.35 | 12.0 | -0.04 | 10.28 | $10.52^{\pm 0.16}$ | $9.92^{\pm 0.23}$ | $3.13^{\pm 1.02}$ | $0.27^{\pm 0.18}$ | $155^{\pm 31}$ | $21^{\pm 5}$ | 11.8 | -0.06 | $8.19^{\pm 0.99}$ |
| (54) | GS4 43501 | 1.61 | 0.55 | 22.0 | 0.07 | 10.61 | $10.88^{\pm 0.10}$ | $10.29^{\pm 0.18}$ | $4.88^{\pm 1.14}$ | $0.31^{\pm 0.08}$ | $240^{\pm 48}$ | $49^{\pm 5}$ | 12.04 | -0.15 | $8.44^{\pm 0.54}$ |
| (55) | GS4 14152 | 1.62 | 0.55 | 11.65 | 0.13 | 11.3 | $11.52^{\pm 0.09}$ | $10.88^{\pm 0.16}$ | $6.93^{\pm 1.26}$ | $0.14^{\pm 0.09}$ | $350^{\pm 70}$ | $40^{\pm 8}$ | 12.8 | 0.04 | $8.26^{\pm 0.93}$ |
| (56) | GS4 16814 | 1.62 | 0.55 | 6.0 | 0.52 | 11.19 | $10.89^{\pm 0.06}$ | $10.73^{\pm 0.08}$ | $5.00^{\pm 1.39}$ | $0.06^{\pm 0.06}$ | $227^{\pm 45}$ | $47^{\pm 7}$ | 12.07 | 0.58 | $7.63^{\pm 0.86}$ |
| (57) | U4 34480 | 1.65 | 0.53 | 20.0 | 0.29 | 10.18 | $10.25^{\pm 0.12}$ | $9.55^{\pm 0.24}$ | $1.57^{\pm 0.62}$ | $0.27^{\pm 0.12}$ | $193^{\pm 39}$ | $29^{\pm 6}$ | 11.57 | 0.8 | $8.68^{\pm 0.62}$ |
| (58) | COS4 05606 | 1.66 | 0.62 | 7.8 | 0.59 | 9.9 | $10.63^{\pm 0.16}$ | $10.33^{\pm 0.18}$ | $4.29^{\pm 1.39}$ | $0.35^{\pm 0.12}$ | $205^{\pm 41}$ | $36^{\pm 9}$ | 11.7 | -0.01 | $8.40^{\pm 0.79}$ |
| (59) | U4 19708 | 1.66 | 0.52 | 7.0 | 0.29 | 11.26 | $10.84^{\pm 0.08}$ | $10.61^{\pm 0.11}$ | $5.88^{\pm 1.55}$ | $0.32^{\pm 0.10}$ | $219^{\pm 44}$ | $34^{\pm 4}$ | 12.06 | 0.29 | $8.28^{\pm 0.51}$ |
| (60) | GS3 15675 | 2.0 | 0.55 | 14.9 | 0.56 | 10.89 | $10.76^{\pm 0.17}$ | $9.76^{\pm 0.25}$ | $7.39^{\pm 1.35}$ | $0.56^{\pm 0.20}$ | $206^{\pm 41}$ | $82^{\pm 6}$ | 11.85 | 0.28 | $8.37^{\pm 0.64}$ |
| (61) | zC 412369 | 2.02 | 0.45 | 6.0 | 0.29 | 10.34 | $10.64^{\pm 0.19}$ | $8.94^{\pm 1.02}$ | $4.38^{\pm 1.31}$ | $0.45^{\pm 0.21}$ | $197^{\pm 39}$ | $94^{\pm 8}$ | 11.78 | 0.19 | $8.47^{\pm 0.96}$ |
| (62) | K20 ID9 | 2.04 | 0.45 | 6.0 | 0.03 | 10.65 | $10.91^{\pm 0.13}$ | $10.39^{\pm 0.18}$ | $7.11^{\pm 1.40}$ | $0.45^{\pm 0.15}$ | $234^{\pm 47}$ | $24^{\pm 8}$ | 12.02 | -0.29 | $8.41^{\pm 0.56}$ |
| (63) | COS4 10347 | 2.06 | 0.44 | 14.9 | 0.2 | 10.76 | $10.81^{\pm 0.11}$ | $10.51^{\pm 0.14}$ | $5.19^{\pm 1.46}$ | $0.21^{\pm 0.18}$ | $220^{\pm 44}$ | $46^{\pm 7}$ | 11.95 | 0.2 | $8.16^{\pm 1.03}$ |
| (64) | COS4 13174 | 2.1 | 0.46 | 18.8 | 0.15 | 11.03 | $11.28^{\pm 0.07}$ | $10.93^{\pm 0.10}$ | $6.99^{\pm 0.94}$ | $0.16^{\pm 0.10}$ | $268^{\pm 54}$ | $41^{\pm 8}$ | 12.38 | 0.05 | $8.10^{\pm 0.69}$ |
| (65) | GS4 01529 | 2.12 | 0.63 | 8.9 | 0.95 | 10.55 | $11.33^{\pm 0.12}$ | $9.63^{\pm 1.46}$ | $4.50^{\pm 1.06}$ | $0.11^{\pm 0.10}$ | $313^{\pm 61}$ | $64^{\pm 10}$ | 12.11 | 1.01 | $8.24^{\pm 1.04}$ |
| (66) | CDFS 2954 | 2.13 | 0.35 | 6.0 | 0.15 | 10.6 | $10.97^{\pm 0.20}$ | $9.45^{\pm 0.79}$ | $4.29^{\pm 1.07}$ | $0.55^{\pm 0.21}$ | $324^{\pm 65}$ | $114^{\pm 9}$ | 12.05 | 0.28 | $9.00^{\pm 0.79}$ |
| (67) | zC 405501 | 2.15 | 0.3 | 7.8 | 0.6 | 9.92 | $10.68^{\pm 0.14}$ | $9.59^{\pm 0.27}$ | $7.13^{\pm 1.03}$ | $0.21^{\pm 0.14}$ | $89^{\pm 18}$ | $61^{\pm 4}$ | 11.72 | -0.42 | $7.23^{\pm 0.60}$ |
| (68) | COS4 08775 | 2.16 | 0.47 | 10.8 | -0.3 | 10.35 | $10.46^{\pm 0.14}$ | $9.53^{\pm 0.32}$ | $4.24^{\pm 1.37}$ | $0.17^{\pm 0.13}$ | $129^{\pm 26}$ | $48^{\pm 5}$ | 11.75 | -0.32 | $7.70^{\pm 0.97}$ |
| (69) | COS4 05094 | 2.17 | 0.49 | 10.7 | 0.01 | 10.4 | $10.77^{\pm 0.12}$ | $10.55^{\pm 0.14}$ | $4.95^{\pm 1.05}$ | $0.30^{\pm 0.16}$ | $223^{\pm 45}$ | $50^{\pm 7}$ | 11.91 | -0.09 | $8.34^{\pm 0.84}$ |
| (70) | SSA22 MD41 | 2.17 | 0.55 | 7.0 | 0.6 | 9.86 | $10.46^{\pm 0.32}$ | $9.15^{\pm 0.52}$ | $7.61^{\pm 1.31}$ | $0.74^{\pm 0.20}$ | $190^{\pm 38}$ | $73^{\pm 6}$ | 11.51 | -0.15 | $8.42^{\pm 0.49}$ |
| (71) | BX389 | 2.17 | 0.45 | 9.0 | 0.1 | 10.6 | $11.05^{\pm 0.18}$ | $10.52^{\pm 0.22}$ | $6.95^{\pm 1.01}$ | $0.56^{\pm 0.13}$ | $302^{\pm 60}$ | $79^{\pm 6}$ | 12.1 | -0.18 | $8.73^{\pm 0.51}$ |
| (72) | U4 36568 | 2.18 | 0.52 | 18.8 | 0.34 | 11.05 | $11.19^{\pm 0.11}$ | $10.97^{\pm 0.14}$ | $3.71^{\pm 1.38}$ | $0.03^{\pm 0.06}$ | $342^{\pm 68}$ | $77^{\pm 6}$ | 12.28 | 0.84 | $7.90^{\pm 1.06}$ |
| (73) | zC 407302 | 2.18 | 0.27 | 19.0 | 0.77 | 10.39 | $10.74^{\pm 0.09}$ | $10.44^{\pm 0.12}$ | $4.64^{\pm 1.16}$ | $0.40^{\pm 0.09}$ | $225^{\pm 45}$ | $43^{\pm 5}$ | 11.74 | 0.7 | $8.51^{\pm 0.50}$ |
| (74) | GS3 24273 | 2.19 | 0.57 | 15.7 | 0.31 | 11.0 | $10.84^{\pm 0.08}$ | $10.74^{\pm 0.09}$ | $7.69^{\pm 1.40}$ | $0.27^{\pm 0.10}$ | $211^{\pm 42}$ | $23^{\pm 7}$ | 11.96 | 0.16 | $8.07^{\pm 0.51}$ |
| (75) | U4 20704 | 2.19 | 0.4 | 18.9 | -0.09 | 11.16 | $11.39^{\pm 0.10}$ | $10.71^{\pm 0.16}$ | $8.89^{\pm 1.41}$ | $0.19^{\pm 0.13}$ | $234^{\pm 47}$ | $80^{\pm 7}$ | 12.4 | -0.02 | $7.93^{\pm 0.81}$ |
| (76) | COS4 05433 | 2.19 | 0.55 | 9.4 | -0.12 | 10.6 | $10.80^{\pm 0.11}$ | $10.50^{\pm 0.14}$ | $5.40^{\pm 1.27}$ | $0.22^{\pm 0.15}$ | $206^{\pm 41}$ | $40^{\pm 6}$ | 11.95 | -0.02 | $8.10^{\pm 0.86}$ |
| (77) | zC 406690 | 2.2 | 0.3 | 10.0 | 0.55 | 10.62 | $11.05^{\pm 0.04}$ | $11.00^{\pm 0.05}$ | $4.82^{\pm 1.21}$ | $0.04^{\pm 0.06}$ | $276^{\pm 55}$ | $72^{\pm 4}$ | 11.92 | 0.61 | $7.64^{\pm 0.85}$ |
| (78) | BX610 | 2.21 | 0.6 | 38.0 | -0.34 | 11.0 | $11.04^{\pm 0.11}$ | $10.64^{\pm 0.16}$ | $5.24^{\pm 1.41}$ | $0.38^{\pm 0.10}$ | $307^{\pm 61}$ | $77^{\pm 5}$ | 12.04 | -0.16 | $8.70^{\pm 0.61}$ |
| (79) | K20 ID8 | 2.22 | 0.59 | 10.7 | -0.28 | 10.5 | $10.70^{\pm 0.10}$ | $10.40^{\pm 0.13}$ | $5.67^{\pm 1.49}$ | $0.16^{\pm 0.12}$ | $168^{\pm 34}$ | $45^{\pm 7}$ | 11.91 | -0.35 | $7.77^{\pm 0.81}$ |
| (80) | U3 10584 | 2.22 | 0.61 | 7.4 | 0.03 | 10.5 | $10.61^{\pm 0.15}$ | $10.31^{\pm 0.17}$ | $6.17^{\pm 1.44}$ | $0.40^{\pm 0.18}$ | $181^{\pm 36}$ | $57^{\pm 7}$ | 11.82 | -0.26 | $8.20^{\pm 0.73}$ |

| | Galaxy | $z$ | FWHM ["] | $T_{int}$ [hrs] | $\delta \log(SFR/SFR(MS))$ | $\log(M_\star/M_\odot)$ | $\log(M_{baryon}/M_\odot)$ | $\log(M_{bulge}/M_\odot)$ | $R_e/kpc$ | $f_{DM}(<R_e)$ | $V_c(R_e)$ [km/s] | $\sigma_0$ [km/s] | $\log(M_{vir}^{SMHM}/M_\odot)$ | $\log(\Sigma_{SFR}/M_\odot yr^{-1}kpc^{-2})$ | $\log \Sigma_{DM}(<R_e)/M_\odot kpc^{-2}$ |
|---|---|---|---|---|---|---|---|---|---|---|---|---|---|---|---|
| (81) | K20 ID7 | 2.22 | 0.4 | 33.0 | 0.3 | 10.37 | $11.04^{\pm 0.23}$ | $9.56^{\pm 0.64}$ | $7.50^{\pm 1.77}$ | $0.65^{\pm 0.17}$ | $282^{\pm 56}$ | $72^{\pm 5}$ | 12.02 | -0.24 | $8.71^{\pm 0.54}$ |
| (82) | U4 13704 | 2.22 | 0.49 | 10.7 | -0.14 | 10.39 | $10.62^{\pm 0.19}$ | $9.62^{\pm 0.34}$ | $3.59^{\pm 1.08}$ | $0.41^{\pm 0.21}$ | $193^{\pm 39}$ | $39^{\pm 8}$ | 11.83 | 0.05 | $8.49^{\pm 0.82}$ |
| (83) | K20 ID5 | 2.22 | 0.29 | 38.0 | 0.26 | 11.0 | $11.18^{\pm 0.08}$ | $11.17^{\pm 0.08}$ | $2.48^{\pm 1.09}$ | $0.07^{\pm 0.08}$ | $174^{\pm 35}$ | $100^{\pm 16}$ | 12.27 | 1.26 | $7.82^{\pm 0.86}$ |
| (84) | K20 ID6 | 2.24 | 0.4 | 25.7 | 0.17 | 10.43 | $10.66^{\pm 0.12}$ | $10.13^{\pm 0.16}$ | $5.19^{\pm 1.10}$ | $0.27^{\pm 0.17}$ | $154^{\pm 31}$ | $62^{\pm 5}$ | 11.8 | 0.07 | $7.96^{\pm 0.85}$ |
| (85) | zC 400569 | 2.24 | 0.25 | 22.0 | 0.19 | 11.08 | $10.95^{\pm 0.07}$ | $10.79^{\pm 0.10}$ | $4.31^{\pm 1.31}$ | $0.16^{\pm 0.09}$ | $241^{\pm 48}$ | $58^{\pm 5}$ | 12.1 | 0.61 | $8.21^{\pm 0.88}$ |
| (86) | BX482 | 2.26 | 0.25 | 18.4 | 0.2 | 10.26 | $10.97^{\pm 0.17}$ | $9.27^{\pm 0.61}$ | $5.98^{\pm 1.03}$ | $0.60^{\pm 0.15}$ | $285^{\pm 57}$ | $71^{\pm 4}$ | 11.98 | -0.15 | $8.78^{\pm 0.52}$ |
| (87) | J0901+1814 | 2.26 | 0.08 | 28.0 | 0.62 | 10.96 | $11.09^{\pm 0.07}$ | $10.09^{\pm 0.28}$ | $4.33^{\pm 0.85}$ | $0.06^{\pm 0.06}$ | $209^{\pm 42}$ | $55^{\pm 3}$ | 12.17 | 0.67 | $7.64^{\pm 0.66}$ |
| (88) | zC 405226 | 2.29 | 0.3 | 14.5 | 0.62 | 9.96 | $10.19^{\pm 0.14}$ | $9.99^{\pm 0.15}$ | $6.15^{\pm 1.37}$ | $0.69^{\pm 0.10}$ | $167^{\pm 33}$ | $68^{\pm 6}$ | 11.44 | -0.01 | $8.36^{\pm 0.47}$ |
| (89) | COS4 02672 | 2.31 | 0.5 | 14.0 | -0.1 | 10.57 | $10.83^{\pm 0.18}$ | $9.83^{\pm 0.31}$ | $7.11^{\pm 1.32}$ | $0.40^{\pm 0.22}$ | $199^{\pm 40}$ | $62^{\pm 4}$ | 11.95 | -0.34 | $8.21^{\pm 0.79}$ |
| (90) | U3 06856 | 2.31 | 0.48 | 15.3 | 0.19 | 10.48 | $10.56^{\pm 0.14}$ | $9.56^{\pm 0.36}$ | $2.48^{\pm 0.86}$ | $0.26^{\pm 0.18}$ | $195^{\pm 39}$ | $53^{\pm 12}$ | 11.77 | 0.81 | $8.47^{\pm 1.02}$ |
| (91) | D3a 15504 | 2.38 | 0.35 | 47.0 | -0.06 | 11.04 | $11.11^{\pm 0.08}$ | $10.59^{\pm 0.12}$ | $6.61^{\pm 1.23}$ | $0.24^{\pm 0.09}$ | $234^{\pm 47}$ | $69^{\pm 4}$ | 12.22 | 0.03 | $8.17^{\pm 0.50}$ |
| (92) | D3a 6004 | 2.39 | 0.4 | 23.0 | -0.05 | 11.5 | $11.47^{\pm 0.08}$ | $11.13^{\pm 0.11}$ | $5.52^{\pm 1.20}$ | $0.10^{\pm 0.11}$ | $359^{\pm 72}$ | $65^{\pm 5}$ | 12.58 | 0.57 | $8.23^{\pm 1.07}$ |
| (93) | zC 400528 | 2.39 | 0.32 | 6.0 | -0.05 | 11.04 | $10.87^{\pm 0.06}$ | $10.82^{\pm 0.08}$ | $1.75^{\pm 0.78}$ | $0.03^{\pm 0.10}$ | $308^{\pm 62}$ | $74^{\pm 11}$ | 12.09 | 1.19 | $8.02^{\pm 1.09}$ |
| (94) | GS4 37124 | 2.43 | 0.43 | 16.6 | 0.28 | 10.59 | $10.73^{\pm 0.10}$ | $10.58^{\pm 0.12}$ | $4.14^{\pm 1.15}$ | $0.39^{\pm 0.13}$ | $253^{\pm 51}$ | $66^{\pm 9}$ | 11.86 | 0.56 | $8.66^{\pm 0.61}$ |
| (95) | COS4 06079 | 2.44 | 0.49 | 9.4 | 0.01 | 10.57 | $11.03^{\pm 0.12}$ | $10.59^{\pm 0.16}$ | $5.26^{\pm 1.37}$ | $0.07^{\pm 0.13}$ | $239^{\pm 48}$ | $36^{\pm 6}$ | 12.09 | 0.07 | $7.76^{\pm 0.94}$ |
| (96) | zC 413597 | 2.45 | 0.35 | 5.8 | 0.56 | 9.87 | $10.13^{\pm 0.11}$ | $9.97^{\pm 0.13}$ | $2.02^{\pm 1.34}$ | $0.09^{\pm 0.11}$ | $136^{\pm 26}$ | $40^{\pm 9}$ | 11.48 | 0.82 | $7.79^{\pm 1.23}$ |
| (97) | GS4 42930 | 2.45 | 0.2 | 19.65 | -0.09 | 10.33 | $10.39^{\pm 0.09}$ | $10.09^{\pm 0.12}$ | $3.67^{\pm 0.88}$ | $0.28^{\pm 0.13}$ | $154^{\pm 31}$ | $50^{\pm 4}$ | 11.68 | 0.22 | $8.12^{\pm 0.72}$ |
| (98) | zC 410041 | 2.45 | 0.37 | 15.2 | 0.51 | 9.66 | $9.88^{\pm 0.19}$ | $9.48^{\pm 0.21}$ | $4.45^{\pm 1.49}$ | $0.63^{\pm 0.19}$ | $124^{\pm 25}$ | $56^{\pm 5}$ | 11.39 | -0.32 | $8.21^{\pm 0.74}$ |
| (99) | COS4 19753 | 2.47 | 0.48 | 8.2 | 0.16 | 10.51 | $10.59^{\pm 0.12}$ | $8.59^{\pm 2.14}$ | $4.22^{\pm 1.06}$ | $0.08^{\pm 0.12}$ | $134^{\pm 27}$ | $73^{\pm 7}$ | 11.76 | 0.39 | $7.38^{\pm 0.81}$ |
| (100) | U3 25105 | 2.52 | 0.57 | 7.3 | 0.08 | 11.05 | $11.06^{\pm 0.07}$ | $10.75^{\pm 0.11}$ | $5.72^{\pm 1.45}$ | $0.08^{\pm 0.08}$ | $239^{\pm 48}$ | $53^{\pm 7}$ | 12.16 | 0.36 | $7.77^{\pm 0.95}$ |

# Appendix C

We used each of the methods (A, B or C) discussed in section 3 to individually fit each galaxy of the RC100 set, for which the four primary galaxy parameters were obtained: $f_{DM}(R_e)$, $\log M_{baryon}$, $R_e$, and $\sigma_0$. In total, this accumulates to 400 values for each method considered. The average difference normalized by the combined uncertainty for each of the model parameters is given in table 2. A value of $< 1$ represents a good agreement between fitting methods for that parameter. In Figure 11 we compare the best fit values of each method against another method, for each of the four model parameters. We also note the median uncertainty, the Pearson correlation coefficients and the median absolute difference in each panel. All of the parameters correlate strongly with one another, with Pearson coefficients of $\geq 0.72$ and up to 0.92. Some variation in inferred values is expected, as degeneracy between model parameters can be susceptible to the way the parameter space is sampled (See Price et al. 2021 for further discussion regarding degeneracies).

|         | $\Delta f_{DM}(R_e)$ | $\Delta \log M_{baryon}$ | $\Delta R_e$ | $\Delta \sigma_0$ |
|---------|----------------------|--------------------------|--------------|-------------------|
| $A - B$ | 0.77                 | 0.69                     | 0.47         | 1.23              |
| $A - C$ | 0.59                 | 0.66                     | 0.43         | 1.05              |
| $B - C$ | 0.56                 | 0.69                     | 0.53         | 0.62              |

Table 4: Average difference in inferred values between each set of fitting methods, $\Delta X_{ij} = |X_i - X_j| / \sqrt{\Delta X_i^2 + \Delta X_j^2}$, $i, j$ stand for the three fitting methods (A, B and C), $X_i$ is a given model parameter for a galaxy for the method $i$, and $\Delta X_i$ is the uncertainty in the parameter.

Figure 12 shows the differences in inferred values, $\Delta X(i - j) = X(i) - X(j)$, where $X$ is a model parameter and $i, j$ are two different methods used. Zero difference means a perfect agreement between methods, and the uncertainty is given by the horizontal error bar. Two important distinctions can be made: (i) the difference distributions peak around zero and are symmetric meaning there are no strong systematic biases, and (ii) the spread of the distributions is of the order-of-magnitude of the uncertainty. This makes most discrepancies between fitting procedures a feature of the underlying uncertainties and degeneracies of the parameters instead of a genuine discrepancy. Such degeneracies help reveal the limits of the "allowed" part of the parameter space. Therefore, taking the mean of the three values (A, B & C) gives a clearer estimate of each parameter's value.

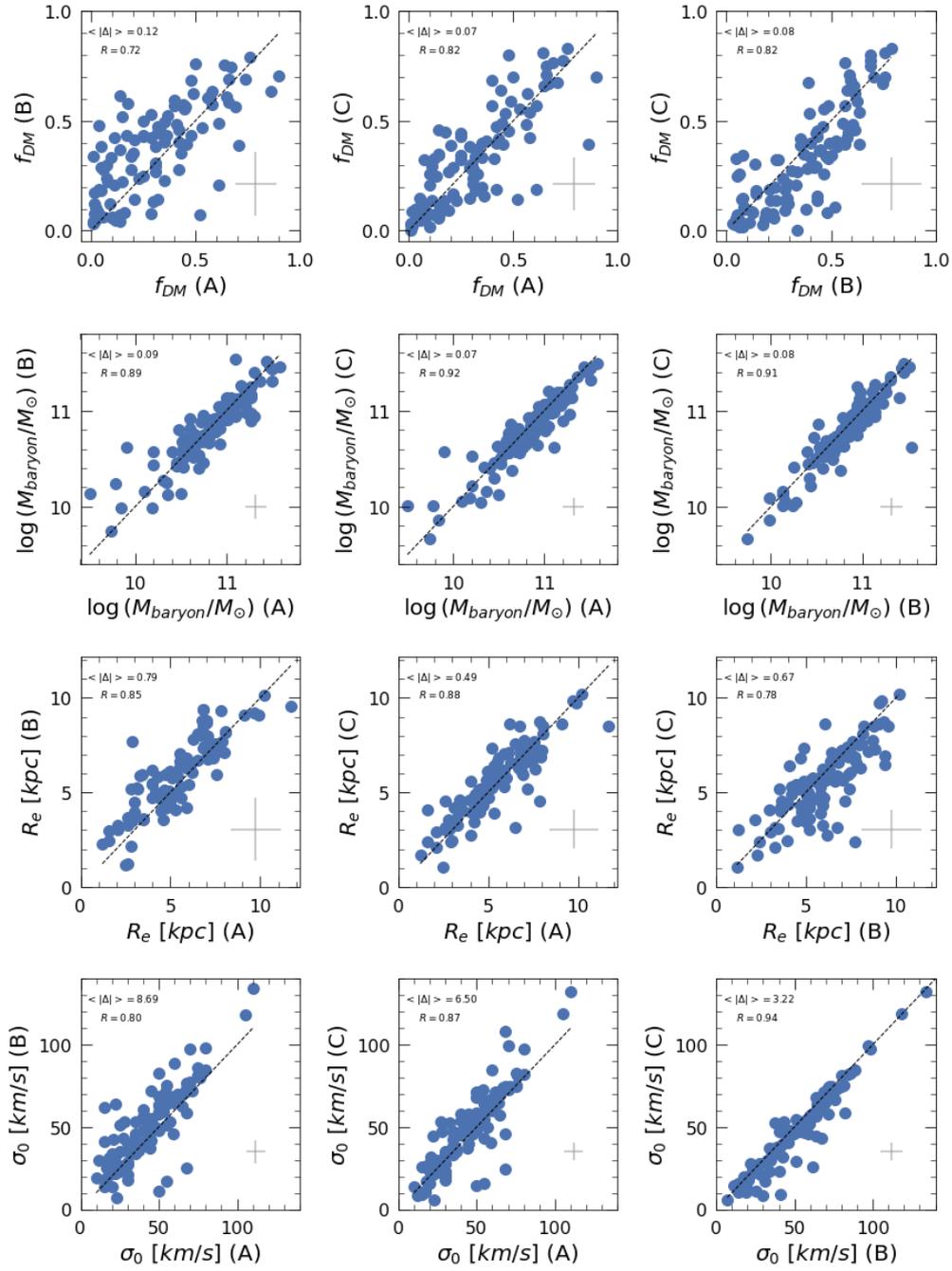

*Figure 11: Parameter* correlations for each of the four primary model parameters: dark matter fractions $f_{DM}(R_e)$, total baryon mass $\log M_{baryon}$, disk effective radius $R_e$, and constant velocity dispersion $\sigma_0$. Each point shows the best fit value for a single galaxy for a combination of two methods: A, B or C. The median error is given in a grey cross and the dashed line shows the one-to-one relation. The absolute median difference is shown in the top left of each panel as $<\Delta>$ (in the same units), and the Pearson correlation coefficient is also *noted.*

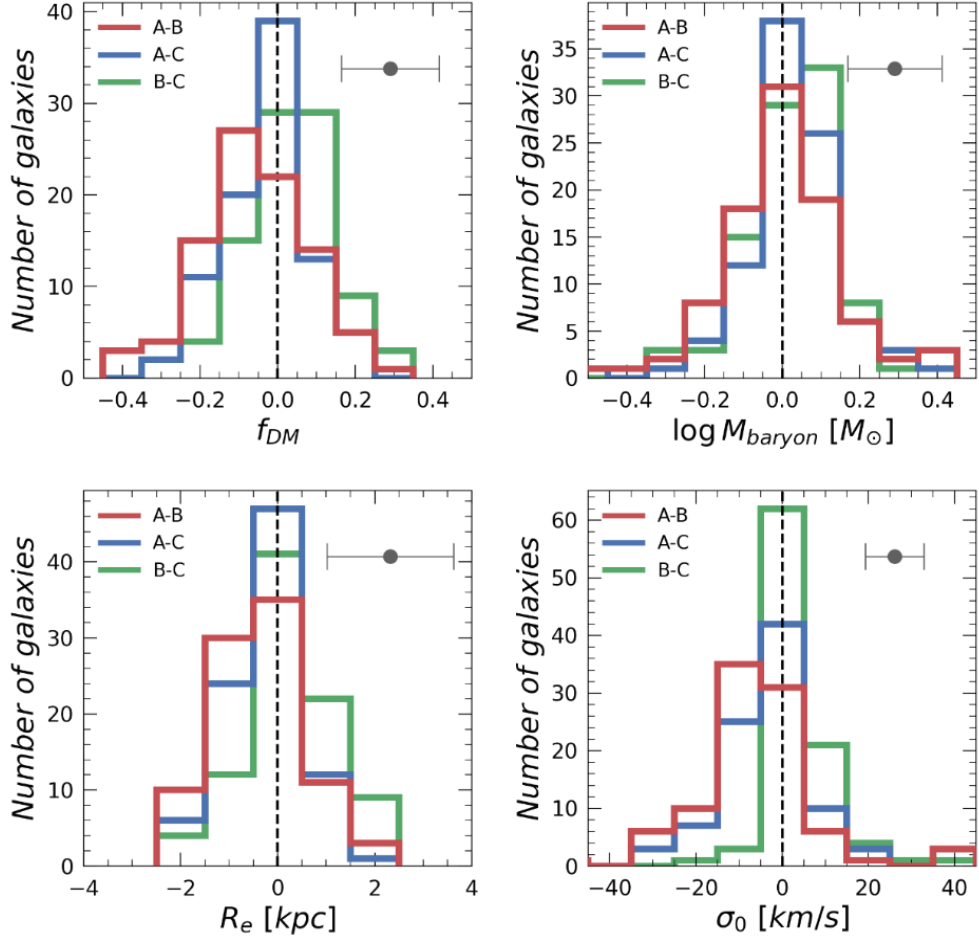

*Figure 12: Histograms of the differences between the best fit galaxy parameters between the three fitting methods: A, B and C. The vertical dashed line represents a one-to-one agreement, and the error bar denotes the typical uncertainty. The majority of the galaxies are close to zero, well within the uncertainties, for every choice of two sets of results.*

# References

van Albada, T. S., & Sancisi, R. 1986, Philos Trans R Soc, 320, 447
Barnabè, M., Dutton, A. A., Marshall, P. J., et al. 2012, MNRAS, 423, 1073
Binney, J., & Tremaine, S. 1987, Galactic Dynamics, Physics Today, Vol. 62
Bland-Hawthorn, J., & Gerhard, O. 2016, ARAA, Vol. 54 (Annual Reviews Inc.), 529
Blumenthal, G. R., Flores, R., & Primack, J. R. 1986, J Chem Inf Model, 301, 27
Boldrini, P. 2021, Galaxies, 10, 5
Bouché, N. F., Bera, S., Krajnovic, D., et al. 2021, A&A, 658, 76
Bovy, J., & Rix, H. W. 2013, ApJ, 779 (Institute of Physics Publishing), 115
Bullock, J. S., Kolatt, T. S., Sigad, Y., et al. 2001, MNRAS, 321, 559
Burkert, A., Förster Schreiber, N. M., Genzel, R., et al. 2016, ApJ, 826, 214
Burkert, A., Genzel, R., Bouché, N., et al. 2010, ApJ, 725, 2324
Cappellari, M., Emsellem, E., Krajnović, D., et al. 2011, MNRAS, 413, 813
Cappellari, M., Scott, N., Alatalo, K., et al. 2013, MNRAS, 432, 1709
Cecil, G., Fogarty, L. M. R., Richards, S., et al. 2016, MNRAS, 456, 1299
Chabrier, G. 2003, Publ Astron Soc Pacific, 115, 763
Chan, T. K., Kereš, D., Oñorbe, J., et al. 2015, MNRAS, 454, 2981
Cooke, L. H., Levy, R. C., Bolatto, A. D., et al. 2022, MNRAS, 000, 1

Courteau, S., & Dutton, A. A. 2015, ApJL, 801, L20
Davies, R. L., Schreiber, N. M. F., Lutz, D., et al. 2020, ApJ, 894, 28
Dekel, A., Freundlich, J., Jiang, F., et al. 2021, MNRAS, 508, 999
Dekel, A., Ishai, G., Dutton, A. A., & Maccio, A. V. 2017, MNRAS, 468, 1005
Dekel, A., & Silk, J. 1986, ApJ, 303 (American Astronomical Society), 39
Dutton, A. A., & Macciò, A. V. 2014, MNRAS, 441, 3359
Dutton, A. A., Treu, T., Brewer, B. J., et al. 2013, MNRAS, 428, 3183
El-Zant, A., Shlosman, I., & Hoffman, Y. 2001, ApJ, 560, 636
Épinat, B., Tasca, L., Amram, P., et al. 2012, A&A, Vol. 539, A92
Faerman, Y., Sternberg, A., & McKee, C. F. 2013, ApJ, 777
Foreman-Mackey, D., Hogg, D. W., Lang, D., & Goodman, J. 2013, Publ Astron Soc Pacific, 125, 306
Förster Schreiber, N. M., Genzel, R., Bouché, N., et al. 2009, ApJ, 706, 1364
Förster Schreiber, N. M., Genzel, R., Lehnert, M. D., et al. 2006, ApJ, 645, 1062
Förster Schreiber, N. M., Renzini, A., Mancini, C., et al. 2018, ApJS, 238, 21
Förster Schreiber, N. M., Übler, H., Davies, R. L., et al. 2019, ApJ, 875, 21
Freundlich, J., Combes, F., Tacconi, L. J., et al. 2019, A&A, Vol. 622, A105
Freundlich, J., Dekel, A., Jiang, F., et al. 2020, MNRAS, 491, 4523
Genzel, R., Förster Schreiber, N. M., Rosario, D., et al. 2014, ApJ, 796, 7
Genzel, R., Price, S. H., Übler, H., et al. 2020, ApJ, 902, 98
Genzel, R., Schreiber, N. M. F., Übler, H., et al. 2017, Nature, 543, 397
Genzel, R., Tacconi, L. J., Kurk, J., et al. 2013, ApJ, 773, 68
Governato, F., Zolotov, A., Pontzen, A., et al. 2012, MNRAS, 422, 1231
Kassin, S. A., de Jong, R. S., & Weiner, B. J. 2006, ApJ, 643, 804
Klypin, A., Kravtsov, A. V., Bullock, J. S., & Primack, J. R. 2001, ApJ, 554, 903
van der Kruit, P. C., & Allen, R. J. 1978, ARAA, 16, 103
Van Der Kruit, P. C., & Freeman, K. C. 2011, ARAA, 49, 301
Kuzio de Naray, R., McGaugh, S. S., de Blok, W. J. G., & Bosma, A. 2006, ApJS, 165, 461
Lang, P., Förster Schreiber, N. M., Genzel, R., et al. 2017, ApJ, 840, 92
Lang, P., Wuyts, S., Somerville, R. S., et al. 2014, ApJ, 788, 11
Lazar, A., Bullock, J. S., Boylan-Kolchin, M., et al. 2020, MNRAS, 497, 2393
Li, Z., Dekel, A., Mandelker, N., Freundlich, J., & François, T. 2022
Lovell, M. R., Pillepich, A., Genel, S., et al. 2018, MNRAS, 481, 1950
Ludlow, A. D., Navarro, J. F., Angulo, R. E., et al. 2014, MNRAS, 441, 378
Madau, P., & Dickinson, M. 2014, ARAA, Vol. 52, 415
Mancini, C., Förster Schreiber, N. M., Renzini, A., et al. 2011, ApJ, 743, 86
Martinsson, T. P. K., Verheijen, M. A. W., Westfall, K. B., et al. 2013, A&A, 557, A131
Mendel, J. T., Beifiori, A., Saglia, R. P., et al. 2020, ApJ, 899, 87
Mo, H. J., Mao, S., & White, S. D. M. 1998, MNRAS, 295, 319
Molina, J., Ibar, E., Smail, I., et al. 2019, MNRAS, 487, 4856
Momcheva, I. G., Brammer, G. B., van Dokkum, P. G., et al. 2016, ApJS, 225, 27
Moore, B., Quinn, T., Governato, F., Stadel, J., & Lake, G. 1999, MNRAS, 310, 1147
Moster, B. P., Naab, T., & White, S. D. M. 2018, MNRAS, 477, 1822
Navarro, J. F., Eke, V. R., & Frenk, C. S. 1996a, MNRAS, 283, 72
Navarro, J. F., Frenk, C. S., & White, S. D. M. 1996b, ApJ, 462, 563
Navarro, J. F., Hayashi, E., Power, C., et al. 2004, MNRAS, 349, 1039
Navarro, J. F., Ludlow, A., Springel, V., et al. 2010, MNRAS, 402, 21
Noordermeer, E. 2008, MNRAS, 385, 1359
Ogiya, G., & Mori, M. 2011, ApJL, 736, 2
Ogiya, G., & Nagai, D. 2022, MNRAS, 000, 1
Oh, S. H., Hunter, D. A., Brinks, E., et al. 2015, Astron J, 149, 1
Power, C., Navarro, J. F., Jenkins, A., et al. 2003, MNRAS, 338, 14
Price, S. H., Kriek, M., Barro, G., et al. 2020, ApJ, 894, 91
Price, S. H., Shimizu, T. T., Genzel, R., et al. 2021, ApJ, 922, 143
Remus, R. S., Dolag, K., Naab, T., et al. 2017, MNRAS, 464, 3742
Schulze, F., Remus, R. S., Dolag, K., et al. 2020, MNRAS, 493, 3778

Simons, R. C., Kassin, S. A., Weiner, B. J., et al. 2017, ApJ, 843, 46
Skelton, R. E., Whitaker, K. E., Momcheva, I. G., et al. 2014, ApJS, 214
De Souza, R. S., Rodrigues, L. F. S., Ishida, E. E. O., & Opher, R. 2011, MNRAS, 415, 2969
Spekkens, K., Giovanelli, R., & Haynes, M. P. 2005, Astron J, 129, 2119
Stott, J. P., Swinbank, A. M., Johnson, H. L., et al. 2016, MNRAS, 457, 1888
Tacchella, S., Lang, P., Carollo, C. M., et al. 2015, ApJ, 802
Tacconi, L. J., Genzel, R., Saintonge, A., et al. 2018, ApJ, 853, 179
Tacconi, L. J., Genzel, R., & Sternberg, A. 2020, ARAA, Vol. 58, 157
Tacconi, L. J., Neri, R., Genzel, R., et al. 2013, ApJ, 768, 74
Tiley, A. L., Swinbank, A. M., Harrison, C. M., et al. 2019, MNRAS, 485, 834
Übler, H., Genel, S., Sternberg, A., et al. 2021, MNRAS, Vol. 500, 4597
Übler, H., Genzel, R., Tacconi, L. J., et al. 2018, ApJ, 854, L24
Übler, H., Genzel, R., Wisnioski, E., et al. 2019, ApJ, 880, 48
Waterval, S., Elgamal, S., Nori, M., et al. 2022, MNRAS
Van Der Wel, A., Franx, M., Van Dokkum, P. G., et al. 2014, ApJ, 788, 28
Wisnioski, E., Förster Schreiber, N. M., Wuyts, S., et al. 2015, ApJ, 799, 209
Wisnioski, E., Schreiber, N. M. F., Fossati, M., et al. 2019, ApJ, 886, 124
Wuyts, S., Förster Schreiber, N. M., Van Der Wel, A., et al. 2011, ApJ, 742
Wuyts, S., Schreiber, N. M. F., Wisnioski, E., et al. 2016, ApJ, 831, 149